\input amstex
\magnification 1200
\TagsOnRight
\def\qed{\ifhmode\unskip\nobreak\fi\ifmmode\ifinner\else
 \hskip5pt\fi\fi\hbox{\hskip5pt\vrule width4pt
 height6pt depth1.5pt\hskip1pt}}
\NoBlackBoxes \baselineskip 19.7 pt
\parskip 5 pt

\centerline {\bf SMALL-ENERGY ANALYSIS FOR}
\vskip -4 pt
\centerline {\bf  THE SELFADJOINT MATRIX SCHR\"ODINGER OPERATOR}
\vskip -4 pt
\centerline {\bf ON THE HALF LINE. II}

\vskip 5 pt
\centerline {Tuncay Aktosun}
\vskip -8 pt
\centerline {Department of Mathematics}
\vskip -8 pt
\centerline {University of Texas at Arlington}
\vskip -8 pt
\centerline {Arlington, TX 76019-0408, USA}
\vskip -8 pt
\centerline {aktosun\@uta.edu}

\centerline {Martin Klaus}
\vskip -8 pt
\centerline {Department of Mathematics}
\vskip -8 pt
\centerline {Virginia Tech}
\vskip -8 pt
\centerline {Blacksburg, VA 24061, USA}
\vskip -8 pt
\centerline {mklaus\@math.vt.edu}

\centerline {Ricardo Weder\plainfootnote{$^\dagger$}
{Fellow Sistema Nacional de Investigadores}}
\vskip -8 pt
\centerline {Departamento de F\'{\i}sica Matem\'atica}
\vskip -8 pt
\centerline {Instituto de Investigaciones en
Matem\'aticas Aplicadas y en Sistemas}
\vskip -8 pt
\centerline {Universidad Nacional Aut\'onoma de M\'exico}

\vskip -8 pt
\centerline {Apartado Postal 20-726, IIMAS-UNAM, M\'exico DF 01000, M\'exico}
\vskip -8 pt \centerline {weder\@unam.mx}

\noindent {\bf Abstract}: The matrix Schr\"odinger equation
with a selfadjoint matrix potential is considered on the half
line with the most general selfadjoint boundary condition at
the origin. When the matrix potential is integrable and has a
second moment, it is shown that the corresponding scattering
matrix is differentiable at zero energy. An explicit formula is
provided for the derivative of the scattering matrix at zero energy. The
previously established results when the potential
has only the first moment are improved when the second moment exists,
by presenting the small-energy asymptotics for the related
Jost matrix, its inverse, and various other quantities relevant
to the corresponding direct and inverse scattering problems.

\par \noindent {\bf Mathematics Subject Classification (2010):}
34L25 34L40  81U05 81Uxx
\vskip -8 pt
\par\noindent {\bf Keywords:}
matrix Schr\"odinger
equation, selfadjoint boundary condition, Jost matrix,
\vskip -8 pt
small-energy limit, scattering matrix, Jost solution,
quantum wires, quantum graphs
\vskip -8 pt
\par\noindent {\bf Short title:} Half-line matrix Schr\"odinger
equation

\newpage

\noindent {\bf 1. INTRODUCTION}
\vskip 3 pt

Consider the matrix Schr\"odinger equation on the half line
$$-\psi''+V(x)\,\psi=k^2\psi,\qquad x\in{\bold R}^+,\tag 1.1$$
where ${\bold R}^+:=(0,+\infty),$
the prime denotes the $x$-derivative, the potential
$V$ is an $n\times n$ matrix-valued function of $x,$ and the wavefunction
$\psi$ is either a column vector with $n$ components
or a matrix of size $n\times n.$ Here, $n$ is any fixed positive integer.
The matrix potential
$V$ is assumed to be selfadjoint, i.e.
$$V(x)=V(x)^\dagger,\tag 1.2$$
where the dagger denotes the adjoint (complex
conjugate and matrix transpose), and hence
$V$ is not necessarily real valued when $n\ge 2.$
We further assume that $V$ belongs to
class $L^1_2({\bold R}^+).$ By $V\in L^1_j({\bold R}^+)$ we mean that
each entry of the matrix $V$ is Lebesgue measurable on
${\bold R}^+$ and
$$\int_0^\infty
dx\,(1+|x|)^j\,||V(x)||<+\infty,$$
where $||V(x)||$ denotes a
matrix norm.
Since all matrix norms are
equivalent, without loss of generality, we can use the matrix norm defined as
$$||V(x)||:=\max_{l}\sum_{s=1}^n |V_{ls}(x)|,\qquad l=1,\dots,n,$$
with $V_{ls}(x)$ denoting the $(l,s)$-entry of the matrix $V(x).$

We use $\bold R$ for the real axis, $\bold C$ for the complex plane,
${\bold C^+}$ for the open upper-half complex plane,
${\overline{\bold C^+}}$ for ${\bold C^+}\cup {\bold R},$ an overdot for the $k$-derivative,
and $I_n$ for the $n\times n$ identity matrix. By a generic constant $c,$ we mean
a constant that does not necessarily take the same value in each appearance.

We are interested in studying (1.1) under the most general
selfadjoint boundary condition at $x=0$ [5,16-18,21,22], which can be stated as
$$-B^\dagger\psi(0)+A^\dagger\psi'(0)=0,\tag 1.3$$
where the constant $n\times n$ matrices $A$ and $B$ satisfy
$$B^\dagger A=A^\dagger B,\quad
A^\dagger A+B^\dagger B>0,\tag 1.4$$
i.e. $A^\dagger B$ is selfadjoint and
the selfadjoint matrix $(A^\dagger
A+B^\dagger B)$ is positive.

By considering the
Schr\"odinger operator corresponding to (1.1)
with a selfadjoint potential in $L^1_2({\bold R}^+)$
and with the selfadjoint boundary condition (1.3),
our primary goal is to
establish the small-$k$
asymptotics of various quantities related to (1.1) such as certain
scattering solutions, the Jost matrix, the inverse of the Jost
matrix, and the scattering matrix. In [5], which is considered to be part I,
we have provided
the corresponding small-$k$ analysis under the assumption
that $V$ is selfadjoint and belongs to $L^1_1({\bold R}^+).$ In the current paper,
which is considered to be part II and a sequel to [5], we refine those asymptotics
by using the stronger assumption
$V\in L^1_2({\bold R}^+)$ instead of $V\in L^1_1({\bold R}^+).$
Such an analysis is crucial in the study of the corresponding
direct and inverse scattering problems. Without knowing the behavior
of the scattering data at and near $k=0,$ it is not possible to solve
the relevant inverse scattering problem.
For the corresponding large-$k$ asymptotics of the Jost matrix and of the scattering matrix, we refer the reader to [6] and the references therein.

The direct scattering
problem for (1.1) consists of the determination of the scattering matrix and the
bound-state information when the matrix potential $V$ and the
selfadjoint boundary condition (1.3) are known. On the other hand,
the corresponding inverse scattering problem is to recover the potential and
the boundary condition from an appropriate set of scattering
data. We recall that a bound state for the corresponding Schr\"odinger operator
is an $n\times 1$ matrix-valued solution to (1.1),
satisfying the boundary condition (1.3), and having each entry
square integrable in $x\in{\bold R}^+.$

The small-$k$ analysis of the
scattering matrix is
not easy even in the scalar case. For example, in the
full-line scalar case Deift and Trubowitz [10] stated that the
characterization of the scattering data given by Faddeev [12] might not hold and in fact even the
continuity of the scattering matrix was not clear when the real-valued potential belonged to $L^1_1({\bold R}),$
and they introduced the stronger assumption that the potential belonged to $L^1_2({\bold R}).$ The proof of the
continuity of the scattering matrix was given later in the scalar case
when $L^1_1({\bold R});$ for
details we refer the reader to [19,20] and the references therein.
The techniques needed for the small-$k$ analysis especially in the exceptional case,
i.e. when the Jost matrix is not invertible, are very delicate and hence
a careful analysis is needed.

Besides its mathematical significance, our analysis is
also important due to its direct relevance
to applications in quantum mechanical scattering of
particles with internal structures,
scattering on
graphs [7,8,11,13-15,23-26], and quantum wires [21,22].
For example, the problem under study describes $n$ connected very thin quantum wires
forming a one-vertex graph with open ends. A linear boundary condition is imposed at the vertex
and the behavior on each wire is governed by a Schr\"odinger operator. The
corresponding direct and inverse problems have physical
relevance in designing elementary gates in quantum computing and
nanotubes for microscopic
electronic devices, with strings of atoms forming a star-shaped graph. For further
details regarding applications we refer the reader to [21] and the references therein.

Our paper
complements the study by Agranovich and Marchenko [1], where
the inverse scattering problem is analyzed with attention to the behavior
at $k=0,$ but only under the
Dirichlet boundary condition, i.e. when
$A=0$ in (1.3) so that the boundary condition becomes $\psi(0)=0.$ It also complements
the study by Harmer [16-18] where the general selfadjoint
boundary condition equivalent to (1.3) is used to investigate the inverse
problem for (1.1) but without any corresponding small-$k$ analysis. We
refer the reader to [2,20] for the small-$k$ analysis for
the half-line scalar Schr\"odinger equation, to [3,19] for the
full-line scalar Schr\"odinger equation, to [4] for the full-line matrix
Schr\"odinger equation, and to [5] for the half-line matrix
Schr\"odinger equation with the most general selfadjoint
boundary condition at the origin under the weaker assumption $V\in L^1_1({\bold R}^+).$

Our paper is organized as follows.
In Section~2 we introduce three $n\times n$ matrix-valued
solutions to (1.1) and provide their relevant properties
that are needed later on. In Section~3 we obtain the small-$k$ asymptotics of the key
quantity $P(k)$ defined in (3.1), and in Theorem~3.1 we provide the explicit asymptotics as $k\to 0$
in ${\overline{\bold C^+}}$ for $P(k)$ up to $o(k^2).$ In Section~3
we also present a remarkable result for the
small-$k$ asymptotics of $f'(k,x)\,f(k,x)^{-1}$ at any fixed $x\in{\bold R^+}$ for which
$f(0,x)^{-1}$ exists, as well as a similar result for
the small-$k$ asymptotics of $f(k,x)\,f'(k,x)^{-1}$ at any fixed $x\in{\bold R^+}$ for which
$f'(0,x)^{-1}$ exists. The corresponding asymptotic
expansions are remarkable in the sense that they are obtained explicitly up to
$o(k^2)$ even though the small-$k$ asymptotics can only be provided up to $o(k)$
for the individual matrices $f(k,x)$ and $f'(k,x)$ and their inverses.
In Section~4 we introduce the Jost matrix $J(k)$ and the scattering matrix
$S(k).$
In Section~5 we analyze the small-$k$ asymptotics of $J(k),$ $J(k)^{-1},$ and $S(k)$ in the generic case,
i.e. when $J(0)^{-1}$ exists; we prove the differentiability at $k=0$ for those three
matrices, and we provide their explicit expansions up to $o(k)$ as
$k\to 0.$ In Section~6 we present a $2n\times 2n$ matrix that is always
invertible and that is closely related to $J(0);$ the result is remarkable in the
sense that the invertibility holds even when the $n\times n$ matrix $J(0)$ is not invertible. In Section~7 we analyze the small-$k$ asymptotics
in the exceptional case, i.e. when $J(0)^{-1}$ does not exist, and we show that, when the potential
$V$ in (1.1) is selfadjoint and belongs to $L^1_2({\bold R^+}),$ the corresponding matrices
$J(k),$ $k\,J(k)^{-1},$ and $S(k)$
are differentiable
at $k=0;$ we also provide the explicit asymptotics up to $o(k)$ for those three
 matrices as $k\to 0.$
Finally, in Section~8 we illustrate our results on the small-$k$ asymptotics
of the Jost matrix, the inverse of the Jost matrix, and the scattering matrix by presenting an explicit example.

\vskip 10 pt
\noindent {\bf 2. PROPERTIES OF SOLUTIONS}
\vskip 3 pt

In this section we introduce three $n\times n$ matrix-valued
solutions to (1.1) denoted by
$f(k,x),$ $\omega(k,x),$ and $\varphi(k,x),$ respectively, and
we present their relevant properties that are needed later on.

By taking the matrix adjoint of (1.1)
and using the selfadjointness (1.2) of the potential $V,$
we obtain the adjoint Schr\"odinger equation
$$-\phi''+\phi\,V(x)=k^2\phi,
\qquad x\in{\bold R^+}.\tag 2.1$$
If $\psi(k,x)$ is a solution to (1.1) for a real value
of the parameter $k,$ then $\psi(k,x)^\dagger$ and $\psi(-k,x)^\dagger$ are
solutions to (2.1). Furthermore,
if $\psi(k,x)$ has an analytic extension from $k\in{\bold R}$ to
$k\in{\bold C^+},$ then $\psi(-k^*,x)^\dagger$ also has
an analytic extension from $k\in{\bold R}$ to
$k\in{\bold C^+},$ where we us an asterisk to denote complex conjugation.

Let $[F;G]:=FG'-F'G$ denote the Wronskian.
It is known [4,5]
that for any two $n\times n$ matrix-valued solutions $\psi(k,x)$ and
$\Psi(k,x)$ to (1.1),
the Wronskians
$[\Psi(k,x)^\dagger;\psi(k,x)]$ for real $k$-values is independent of
$x.$
In case $\psi(k,x)$ and $\Psi(k,x)$ have analytic extensions from $k\in{\bold R}$ to
$k\in{\bold C^+},$ the Wronskian
$[\Psi(-k^*,x)^\dagger;\psi(k,x)]$ is independent of $x.$

The Jost solution to
(1.1) is the $n\times n$ matrix-valued solution
satisfying, for each fixed $k\in{\overline{\bold C^+}}\setminus\{0\},$ the spatial asymptotics
$$f(k,x)=e^{ikx}[I_n+o(1)],\quad
f'(k,x)=ik\,e^{ikx}[I_n+o(1)],\qquad x\to+\infty.\tag 2.2$$
 From (1.1) and (2.2) it follows that
$$f(k,x)=e^{ikx}I_n+\displaystyle\frac{1}{k}\int_x^\infty dy\,[\sin k(y-x)]\, V(y)\,f(k,y).\tag 2.3$$
By letting $k\to 0$ in (2.3) we obtain
$$f(0,x)=I_n+\int_x^\infty dy\,(y-x)\, V(y)\,f(0,y).\tag 2.4$$
It can be verified that $f(0,x)$ satisfies
the $n\times n$ matrix-valued zero-energy Schr\"odinger equation
$$-\psi''+V(x)\,\psi=0,\qquad x\in{\bold R}^+,\tag 2.5$$
which is obtained from (1.1) by setting $k=0$ there, as well as the
asymptotic conditions
$$f(0,x)=I_n+o(1),\qquad f'(0,x)=o(1),\qquad x\to+\infty,\tag 2.6$$
which are obtained from (1.1) and (2.2), respectively, by letting $k\to 0.$

Some relevant properties associated with $f(k,x)$ and
$f(0,x)$ are stated in the following two propositions.

\noindent {\bf Proposition 2.1} {\it Assume that the potential $V$ is selfadjoint
and belongs to $L^1_2({\bold R}^+),$ and let $f(k,x)$ be the
corresponding Jost solution to (1.1) appearing in (2.2). Then:}

\item{(a)} {\it For each fixed
$x\in[0,+\infty),$ the quantities $f(k,x)$ and
$f'(k,x)$ are analytic in $k\in{\bold C^+},$ continuous in $k\in{\overline{\bold C^+}},$ and we have}
$$f(k,x)=f(0,x)+k\,\dot f(0,x)+o(k),\qquad k\to 0 \ \text{in} \ {\overline{\bold C^+}},
\tag 2.7$$
$$f'(k,x)=f'(0,x)+k\,\dot f'(0,x)+o(k),\qquad k\to 0 \ \text{in} \ {\overline{\bold C^+}},
\tag 2.8$$
{\it where we recall that an overdot denotes the $k$-derivative.}

\item{(b)} {\it For each fixed $k\in{\overline{\bold C^+}},$ the matrix-valued functions
$f(k,x),$ $f'(k,x),$ $\dot f(k,x),$
and $\dot f'(k,x)$ are continuous
in $x\in[0,+\infty).$}

\item{(c)} {\it For each fixed $k\in{\overline{\bold C^+}}\setminus\{0\},$ the spatial asymptotics in
(2.2) can actually be replaced by the sharper estimates}
$$f(k,x)=e^{ikx}\left[I_n+o\left(\displaystyle\frac{1}{x^2}\right)\right],\quad
f'(k,x)=ik\,e^{ikx}\left[I_n+o\left(\displaystyle\frac{1}{x^2}\right)\right],\qquad x\to+\infty,\tag 2.9$$
{\it as a result of $V\in L^1_2({\bold R}^+).$}

\item{(d)} {\it The Jost solution $f(k,x)$ satisfies the Wronskian relations}
$$[f(\pm k,x)^\dagger;f(\pm k,x)]=
\pm 2ikI_n,\qquad k\in{\bold R},\tag 2.10$$
$$[f(-k^*,x)^\dagger;f(k,x)]=
0,\qquad k\in{\overline{\bold C^+}}.\tag 2.11$$

\item{(e)} {\it When $x$ is large, the matrix $f(0,x)$ is invertible.}

\item{(f)} {\it For every $a\in[0,+\infty)$ for which the matrix $f(0,a)$
is invertible, the matrix $f(k,a)^{-1}$ exists when
$k$ is in a neighborhood of $k=0$ in ${\overline{\bold C^+}}$ and we have}
$$f(k,a)^{-1}=f(0,a)^{-1}-k\,f(0,a)^{-1}\,\dot f(0,a)\, f(0,a)^{-1}+o(k),
\qquad k\to 0 \ \text{in} \ {\overline{\bold C^+}}.
\tag 2.12$$
{\it For every $a\in[0,+\infty)$ for which $f'(0,a)$
is invertible, the matrix $f'(k,a)^{-1}$ exists when
$k$ is in a neighborhood of $k=0$ in ${\overline{\bold C^+}},$ and as
$k\to 0$ in ${\overline{\bold C^+}}$ we have}
$$f'(k,a)^{-1}=f'(0,a)^{-1}-k\,f'(0,a)^{-1}\,\dot f'(0,a)\, f'(0,a)^{-1}+o(k).
\tag 2.13$$

\noindent PROOF: The proof of (a) follows from Theorem~2.1 of [4].
The proof of (b) can be given by using iteration on (2.3) and on
the related integral
 equations obtained by taking the derivatives with respect to $x$ and $k.$
Then, by representing
the relevant quantities as uniformly convergent series of continuous functions,
the continuity in $x$ can be established for those relevant quantities.
The statement of (c) can be obtained from Theorem~1.4.1 of [1].
We quote (2.10) and (2.11) given in (d) from (3.18) and (3.19), respectively, of [5].
The invertibility in (e) is a consequence of (b) and (2.6).
The existence of $f(0,a)^{-1},$ the continuity of
$f(k,a)$ at $k=0$ in ${\overline{\bold C^+}},$ and (2.7) yield (2.12), and we establish
(2.13) in a similar way. \qed

The $n\times n$ matrix-valued zero-energy Schr\"odinger equation
given in (2.5) has [4] two linearly independent $n\times n$ matrix solutions, one of which is bounded and the
other is unbounded. As indicated in the following proposition,
a bounded solution to (2.5) is given by $f(0,x)$
and an unbounded solution is given [4] by $\dot f(0,x).$ Furthermore,
the $2n$ columns of $f(0,x)$ and $\dot f(0,x)$ form $2n$ linearly
independent $n\times 1$ matrix-valued solutions to (2.5).

\noindent {\bf Proposition 2.2} {\it Assume that the potential $V$ is selfadjoint
and belongs to $L^1_2({\bold R}^+),$ and let $f(k,x)$ be the
corresponding Jost solution to (1.1) appearing in (2.2). Then:}

\item{(a)} {\it The zero-energy Jost solution $f(0,x)$ appearing
in (2.6) is a bounded
solution to (2.5), and in fact it satisfies the sharper spatial asymptotics}
$$f(0,x)=I_n+o\left(\displaystyle\frac{1}{x}\right),\quad f'(0,x)=o\left(\displaystyle\frac{1}{x^2}\right),\qquad x\to+\infty.\tag 2.14$$

\item{(b)} {\it We have $f(0,x)-I_n\in L^1({\bold R^+})$ under the weaker
assumption $V\in L_1^1({\bold R^+}).$}

\item{(c)} {\it For each fixed $x\in{\bold R^+},$ we have}
$$\int_x^\infty dy\,V(y)\,f(0,y)=-f'(0,x),\tag 2.15$$
$$\int_x^\infty dy\,y\,V(y)\,f(0,y)=-I_n+f(0,x)-x\,f'(0,x),\tag 2.16$$
$$\int_x^\infty dy\,y^2 \,V(y)\,f(0,y)=-x^2\,f'(0,x)
+2x[f(0,x)-I_n]+2
\int_x^\infty dy\,[f(0,y)-I_n].\tag 2.17$$

\item{(d)} {\it The quantity $\dot f(0,x)$ is an unbounded
solution to (2.5) and it satisfies the spatial asymptotics}
$$\dot f(0,x)=ix \left[I_n+o\left(\displaystyle\frac{1}{x}\right)\right],\quad \dot f'(0,x)=iI_n +o\left(\displaystyle\frac{1}{x}\right) ,\qquad x\to+\infty.\tag 2.18$$

\item{(e)} {\it For each $x\in{\bold R^+},$ we have}
$$\bmatrix f(0,x)& \dot f(0,x)\\
\noalign{\medskip}
f'(0,x)& \dot f'(0,x)\endbmatrix^{-1}=i\bmatrix \dot f'(0,x)^\dagger& -\dot f(0,x)^\dagger\\
\noalign{\medskip}
f'(0,x)^\dagger& -f(0,x)^\dagger\endbmatrix .\tag 2.19$$

\noindent PROOF: We already know (a), (b), and (d) from the Appendix of [4].
We obtain (2.19) in (e) from
$$\bmatrix f(0,x)& \dot f(0,x)\\
\noalign{\medskip}
f'(0,x)& \dot f'(0,x)\endbmatrix\bmatrix \dot f'(0,x)^\dagger& -\dot f(0,x)^\dagger\\
\noalign{\medskip}
f'(0,x)^\dagger& -f(0,x)^\dagger\endbmatrix =\bmatrix -iI_n&0\\
\noalign{\medskip}
0&-i I_n\endbmatrix
,\tag 2.20$$
which is given in the Appendix of [4]. As for (c), we prove (2.15) and (2.16) directly by using (2.4);
with the help of (2.5), (2.14), (b), and integration by parts, we get (2.17). \qed

We define another $n\times n$ matrix-valued solution to (1.1), denoted by
$\omega(k,x),$ which satisfies the initial conditions
$$\omega(k,a)=f(0,a),\quad \omega'(k,a)=f'(0,a),\tag 2.21$$
where $a$ is the constant appearing in Proposition~2.1(f), i.e. $a$ is
a nonnegative constant for which $f(0,a)^{-1}$ exists.
We suppress the dependence on $a$ in our notation for $\omega(k,x).$
Note that $\omega(k,x)$ satisfies [5]
$$\aligned\omega(k,x)=&f(0,a)\,\cos k(x-a)+f'(0,a)\,\displaystyle\frac{\sin k(x-a)}{k}
\\
&+\displaystyle\frac{1}{k}\int_a^x dy\,[\sin k(x-y)]\,V(y)\,\omega(k,y).
\endaligned\tag 2.22$$
 From (2.22), by letting $k\to 0,$ we obtain
$$\omega(0,x)=f(0,a)+f'(0,a)\,(x-a)
+\int_a^x dy\,(x-y)\,V(y)\,\omega(0,y).\tag 2.23$$
Similar to (A19) of [4], we see that $\omega(k,x)$ satisfies
$$\omega(k,x)=\omega(0,x)+ik^2 f(0,x)\int_a^x dy\, \dot f(0,y)^\dagger\, \omega(k,y)
+ik^2 \dot f(0,x)\int_a^x dy\, f(0,y)^\dagger\, \omega(k,y)
.\tag 2.24$$
Note that, with the help of (2.20) and
(2.23), one can directly verify that the right-hand side of (2.24) satisfies (1.1) and (2.21).

Let us write (2.24) as
$$\omega(k,x)=\omega(0,x)+ik^2 \omega_1(x)+
ik^2 \omega_2(k,x),\tag 2.25$$
where we have defined
$$\omega_1(x):=f(0,x)\int_a^x dy\, \dot f(0,y)^\dagger\, \omega(0,y)
+\dot f(0,x)\int_a^x dy\, f(0,y)^\dagger\, \omega(0,y),\tag 2.26$$
$$\aligned \omega_2(k,x):=&f(0,x)\int_a^x dy\, \dot f(0,y)^\dagger\, [\omega(k,y)-\omega(0,y)]\\
&
+\dot f(0,x)\int_a^x dy\, f(0,y)^\dagger\, [\omega(k,y)-\omega(0,y)].
\endaligned\tag 2.27$$

Some relevant properties of $\omega(k,x)$ are provided in the following
proposition.

\noindent {\bf Proposition 2.3} {\it Assume that the potential $V$
appearing in (1.1) is selfadjoint,
and let $\omega(k,x)$ be
the solution to (1.1) appearing in (2.21). Then:}

\item{(a)} {\it When $V$ belongs to $L^1_1({\bold R}^+),$ for each
fixed $x\in{\bold R^+}$ the quantities $\omega(k,x)$ and
$\omega'(k,x)$ are
 entire in $k,$ and we have}
$$\omega(0,x)=f(0,x),\qquad x\in{\bold R}^+,\tag 2.28$$
$$||\omega(k,x)-\omega(0,x)||\le c\left(\displaystyle\frac{|k|(x-a)}
{1+|k|(x-a)}\right)^2 e^{(\text{Im}[k])(x-a)},
\qquad k\in{\overline{\bold C^+}},\quad x\ge a,
\tag 2.29$$
{\it where $c$ is a generic constant.}

\item{(b)} {\it When $V$ belongs to $L^1_2({\bold R}^+),$
the quantity $\omega_2(k,x)$ defined in (2.27) satisfies}
$$||\omega_2(k,x)||\le c\,(1+x^2)\left(\displaystyle\frac{|k|x}{1+|k|x}\right)^2e^{(\text{Im}[k])(x-a)},
\qquad k\in{\overline{\bold C^+}},\quad x\ge a.
\tag 2.30$$

\item{(c)} {\it When $V$ belongs to $L^1_2({\bold R}^+),$
the quantity $\omega_2(k,x)$ defined in (2.27) satisfies}
$$\int_a^\infty dx\, e^{ikx}\, V(x)
\,\omega_2(k,x)=o(1),
\qquad k\to 0 \ {\text{in}} \ {\overline{\bold C^+}}.
\tag 2.31$$

\item{(d)} {\it When $V$ belongs to $L^1_2({\bold R}^+),$ the quantity $\omega_1(x)$
defined in (2.26) satisfies}
$$||\omega_1(x)||\le c\,(1+x^2),\qquad x\in{\bold R^+},\tag 2.32$$
$$\int_a^\infty dx\, [e^{ikx}-1]\, V(x)
\,\omega_1(x)=o(1),
\qquad k\to 0 \ {\text{in}} \ {\overline{\bold C^+}},
\tag 2.33$$
$$\int_a^\infty dx\, V(x)
\,\omega_1(x)=i\int_a^\infty dx\,[f(0,x)^\dagger f(0,x)-I_n]-i
\int_a^\infty dx\,\,[f(0,x)-I_n].
\tag 2.34$$

\noindent PROOF: The analyticity properties stated in (a) follow from the standard
theory of initial-value problems for ordinary differential equations [9].
 Note that (2.28) follows from (2.21), the fact that $\omega(k,x)$
 is a solution to (1.1), and the uniqueness of a corresponding initial-value problem. The estimate in (2.29)
is obtained in Proposition~5.1 of [5]. Let us now prove (2.30). From Proposition~2.1(b)
we know that
$f(0,x)$ and $\dot f(0,x)$ are continuous in $x\in{\bold R^+},$ and hence from
(2.14) and (2.18) we get
$$||f(0,x)||\le c,\quad ||\dot f(0,x)||\le c\,(1+x),
\qquad x\in{\bold R^+}.\tag 2.35$$
Using (2.29) and (2.35) in (2.27), by exploiting the fact that
$z\mapsto z^2/(1+z)^2$ is an increasing function in $z\in{\bold R^+}$ and
$(1+x)^2\le 2(1+x^2)$ for $x\in{\bold R^+},$ we obtain (2.30).
We get (2.31) by using (2.30) and by exploiting the fact that $V\in L^1_2({\bold R}^+).$
We obtain (2.32) by using (2.28) and (2.35) in (2.26).
Using (2.32) and the fact that
$$|e^{iz}-1|\le \displaystyle\frac{c\,|z|}{1+|z|},\qquad z\in{\overline{\bold C^+}},$$
we get
$$||(e^{ikx}-1)\,V(x)\,\omega_1(x)||\le \displaystyle\frac{c\,|k|\,x}{1+|k|\,x}\,(1+x^2)\,||V(x)||,
\qquad x\in{\bold R^+},\quad k\in{\overline{\bold C^+}}.\tag 2.36$$
Thus, (2.36) implies (2.33) when $V\in L^1_2({\bold R^+}).$
Let us finally prove (2.34). First, we note that the integrals on the right-hand side
in (2.34) exist as a result of
Proposition~2.2(c), the boundedness of $f(0,x)$ expressed in
(2.35), and the decomposition
$$f(0,x)^\dagger \,f(0,x)-I_n=[f(0,x)^\dagger-I_n]\,f(0,x)+
[f(0,x)-I_n].\tag 2.37$$
Using (2.28), let us write (2.26) as
$$\omega_1(x)=f(0,x)\,G_1(x)+\dot f(0,x)\,G_2(x),\tag 2.38$$
where we have defined
$$G_1(x):=\int_a^x  dy\, \dot f(0,y)^\dagger\,f(0,y),\quad
G_2(x):=\int_a^x  dy\, f(0,y)^\dagger\,f(0,y).\tag 2.39$$
With the help of (2.35) we see that
$$G_1(x)=O(x^2),\quad G_2(x)=(x-a)I_n+\int_a^\infty
dy\,[f(0,y)^\dagger f(0,y)-I_n]+o(1),\qquad
x\to +\infty.\tag 2.40$$
Note that (2.14), (2.18), and (2.40) yield
$$f'(0,x)\,G_1(x)=o(1),\qquad
x\to +\infty,\tag 2.41$$
$$\dot f'(0,x)\,G_2(x)=i(x-a)I_n+i\int_a^\infty
dy\,[f(0,y)^\dagger f(0,y)-I_n]+o(1),\qquad
x\to +\infty.\tag 2.42$$
Since $f(0,x)$ and $\dot f(0,x)$ are solutions to (2.5), with the help of (2.38) we obtain
$$\int_a^x dy\, V(y)
\,\omega_1(y)=\int_a^x dy\,[f''(0,y)\,G_1(y)+\dot f''(0,y)\,G_2(y)].
\tag 2.43$$
Using integration by parts in (2.43), with the help of (2.38)-(2.42) and the fact
that $G_1(a)=G_2(a)=0,$ we get
$$\aligned
\int_a^x dy\, V(y)
\,\omega_1(y)=&-
\int_a^x dy\, [f'(0,y)\,\dot f(0,y)^\dagger+\dot f'(0,y)\,f(0,y)^\dagger]\,f(0,y)
\\
&+
i(x-a)I_n+i\int_a^\infty
dy\,[f(0,y)^\dagger f(0,y)-I_n]
+o(1),\qquad
x\to +\infty.\endaligned \tag 2.44$$
Using in (2.44) the equality from the (2,2)-entry in (2.20), we obtain (2.34). \qed

Let us introduce the $n\times n$ matrix-valued solution $\varphi(k,x)$ to (1.1)
satisfying the initial conditions
$$\varphi(k,0)=A,\quad \varphi'(k,0)=B,\tag 2.45$$
where $A$ and $B$ are the constant matrices appearing in (1.3). The relevant properties
of $\varphi(k,x)$ are summarized in the following proposition.

\noindent {\bf Proposition 2.4} {\it Assume that the potential $V$ is selfadjoint
and belongs to $L^1_2({\bold R}^+),$ and let $\varphi(k,x)$ be the
corresponding solution to (1.1) appearing in (2.45). Then:}

\item{(a)} {\it For each fixed
$x\in[0,+\infty),$ the quantities $\varphi(k,x)$ and
$\varphi'(k,x)$ are entire in $k\in\bold C.$}

\item{(b)} {\it For each fixed
$x\in[0,+\infty),$ we have}
$$\varphi(k,x)=\varphi(0,x)+O(k^2),\quad \varphi'(k,x)=\varphi'(0,x)+O(k^2)
,\qquad k\to 0 \ \text{in} \ \bold C.
\tag 2.46$$

\noindent PROOF: The properties listed in (a)
are already known [5] and can be established by using
the integral relation given in (3.7) of [5], namely
$$\varphi(k,x)=A\,\cos kx+B\,\displaystyle\frac{\sin kx}{k}+\displaystyle\frac {1}{k}\int
_0^x dy\,[\sin k(x-y)]\,V(y)\,\varphi(k,y).$$
The proof of (b) can be given by using the integral relation
$$\varphi(k,x)=\varphi(0,x)+ik^2 f(0,x)\int_0^x dy\, \dot f(0,y)^\dagger\, \varphi(k,y)
+ik^2 \dot f(0,x)\int_0^x dy\, f(0,y)^\dagger\, \varphi(k,y)
,$$
which, like (2.24), follows by using the method of variation of parameters. \qed

\vskip 10 pt
\noindent {\bf 3. PRELIMINARIES}
\vskip 3 pt

With the goal of obtaining small-$k$ asymptotics of the Jost
matrix, its inverse, and the scattering matrix, in this section we provide
the small-$k$ analysis for certain quantities related to (1.1).
Our first main result is given in Theorem~3.1 and it provides the
asymptotics up to $o(k^2)$ as $k\to 0$ in ${\overline{\bold C^+}}$
for the key $n\times n$ matrix-valued quantity $P(k)$ defined in (3.1). Our second
main result is given in Theorem~3.2, which establishes the
expansions up to $o(k^2)$ as $k\to 0$ in ${\overline{\bold C^+}}$ for the matrices
$f'(k,x)\,f(k,x)^{-1}$ and $f(k,x)\,f'(k,x)^{-1}.$

As in (5.27) of [5], in terms of a Wronskian let us define
$$P(k):=[\omega(-k^*,x)^\dagger;f(k,x)],\tag 3.1$$
where $f(k,x)$ and $\omega(k,x)$ are the solutions to (1.1) appearing
in (2.2) and (2.21), respectively. Note that $P(k)$ depends on the choice of
$a$ because, as apparent from (2.21), $\omega(k,x)$ depends on $a.$ However, we have suppressed $a$ in our notation of
$P(k).$
The Wronskian in (3.1) is independent of $x,$ and its value can be found by using $x=a$ in (3.1) to obtain
$$P(k)=f(0,a)^\dagger f'(k,a)-f'(0,a)^\dagger f(k,a),\tag 3.2$$
where we have used (2.21).

In the following theorem we state some relevant properties of
$P(k),$ which are needed in determining
the small-$k$ asymptotics of the Jost matrix $J(k)$ defined in (4.1).

\noindent {\bf Theorem 3.1} {\it Assume that
the potential $V$ is selfadjoint
and belongs to $L^1_2({\bold R}^+),$ and let $f(k,x)$ be the Jost solution to (1.1)
satisfying (2.2). Then,
the matrix $P(k)$ given in (3.1) is analytic
in ${\bold C^+},$ continuous in ${\overline{\bold C^+}},$ and satisfies}
$$P(k)=ikI_n+k^2\left(-a I_n+ \int_a^\infty dy\,[f(0,y)^\dagger \,f(0,y)-I_n]
\right)+
o(k^2),\qquad k\to 0 \text{ in }{\overline{\bold C^+}}.\tag 3.3$$

\noindent PROOF: The analyticity in ${\bold C^+}$ and the continuity in ${\overline{\bold C^+}}$
follow from Proposition~2.1(a) and Proposition~2.3(a), namely,
the fact that $f(k,x)$ and $f'(k,x)$ satisfy those properties for each fixed $x\in{\bold R^+}$
and that $\omega(k,x)$ and $\omega'(k,x)$ are entire in $k$ for each fixed $x\in{\bold R^+}.$
Let us note that the convergence of the integral in (3.3) is a result of (2.37) and
Proposition~2.2(b).
In order to establish (3.3), we could try to use the Maclaurin
expansions in $k$ for $\omega(k,x)$ and $\omega'(k,x)$ and
the expansions in (2.3) and (2.8). However, we then only get
$P(k)=ikI_n+o(k)$ in (3.3), which is already valid [5] under the weaker assumption
$V\in L^1_1({\bold R^+}).$ Thus, we have to use more sophisticated techniques to
establish the expansion in (3.3). We proceed as follows.
Differentiating (2.22) we get
$$\aligned \omega'(k,x)=&-k\,f(0,a)\,\sin k(x-a)+f'(0,a)\,\cos k(x-a)
\\
&+\int_a^x dy\,[\cos k(x-y)]\,V(y)\,\omega(k,y).\endaligned\tag 3.4$$
Using (2.22) and (3.4) on the right-hand side of (3.1), with the help of
(1.2) we obtain
$$P(k)=P_1(k,x)+P_2(k,x)+P_3(k,x)+P_4(k,x)+P_5(k,x)+P_6(k,x),\tag 3.5$$
where we have defined
$$P_1(k,x):=[\cos k(x-a)]\,f(0,a)^\dagger f'(k,x),\tag 3.6$$
$$P_2(k,x):=\displaystyle\frac{\sin k(x-a)}{k}\,f'(0,a)^\dagger f'(k,x),\tag 3.7$$
$$P_3(k,x):=k\,[\sin k(x-a)]\,f(0,a)^\dagger f(k,x),\tag 3.8$$
$$P_4(k,x):=-[\cos k(x-a)]\,f'(0,a)^\dagger f(k,x),\tag 3.9$$
$$P_5(k,x):=\int_a^x dy\,\omega(0,y)^\dagger\,V(y)\left[
\displaystyle\frac{\sin k(x-y)}{k}\,f'(k,x)-[\cos k(x-y)]\,f(k,x)\right],\tag 3.10$$
$$\aligned P_6&(k,x):=\\
&\int_a^x dy\,\left[\omega(-k^*,y)^\dagger-\omega(0,y)^\dagger\right] \,V(y)\left[
\displaystyle\frac{\sin k(x-y)}{k}\,f'(k,x)-[\cos k(x-y)]\,f(k,x)\right].
\endaligned\tag 3.11$$
Since the value of $P(k)$ defined in (3.1) is independent of $x,$ we will evaluate
the right-hand side in (3.5) as $x\to+\infty.$ With this goal in mind, we
proceed as follows.
Writing the sine and cosine functions in (3.6)-(3.9) in terms of complex
exponentials $e^{ik(x-a)}$ and $e^{-ik(x-a)}$ and using (2.9), from
(3.6)-(3.9) we obtain
$$\aligned
P_1(k,x)+P_2(k,x)+&P_3(k,x)+P_4(k,x)
\\
&=\left[
ik\,e^{ika}f(0,a)^\dagger-e^{ika}f'(0,a)^\dagger \right]+o(1),\qquad x\to+\infty.
\endaligned \tag 3.12$$
Proceeding in a similar manner, with the help of
(2.9), as $x\to+\infty$ from (3.10) and (3.11) we obtain
$$P_5(k,x)=-\int_a^\infty dy\,\omega(0,y)^\dagger\,V(y)\,e^{iky}+o(1),\qquad x\to+\infty,
\tag 3.13$$
$$P_6(k,x)=-\int_a^\infty dy\,\left[\omega(-k^*,y)^\dagger-\omega(0,y)^\dagger\right] \,V(y)\,e^{iky}+o(1),\qquad x\to+\infty.
\tag 3.14$$
Next we will estimate the integrals in (3.13) and (3.14) further, by working
on (3.13) first.
We decompose the integral in (3.13) as
$$P_5(k,x)=P_7(k)+P_8(k)+o(1),\qquad x\to+\infty,\tag 3.15$$
where we have defined
$$P_7(k):=-\int_a^\infty dy\,\omega(0,y)^\dagger\,V(y)\left[
1+iky-\displaystyle\frac{k^2y^2}{2}\right],$$
$$P_8(k):=-\int_a^\infty dy\,\omega(0,y)^\dagger\,V(y)\left[e^{iky}-
1-iky+\displaystyle\frac{k^2y^2}{2}\right].\tag 3.16$$
With the help of (2.14)-(2.17) and (2.28), we can write $P_7(k)$ as
$$P_7(k)=p_1-ik p_2+ k^2 p_3,\tag 3.17$$
where we have defined
$$p_1:=f'(0,a)^\dagger,$$
$$p_2:=-I_n+f(0,a)^\dagger -a\,f'(0,a)^\dagger,$$
$$p_3:=-\displaystyle \frac 12\,a^2 f'(0,a)^\dagger+a[f(0,a)^\dagger-I_n]+\int_a ^\infty dy\,[f(0,y)^\dagger -I_n].$$
Note that for any $z\in{\overline{\bold C^+}}$ we have
$$\left|e^{iz}-1-iz+\displaystyle\frac{z^2}{2} \right|\le \displaystyle\frac{c\,|z|^3}{1+|z|},
\tag 3.18$$
where $c$ denotes a generic positive
constant independent of the complex number $z.$
Using (3.18) in (3.16), we obtain
$$||P_8(k)||\le c\int_a^\infty dy\,||\omega(0,y)^\dagger||\,||V(y)||\left(\displaystyle\frac{|k|y}{1+|ky|}
\right)\left(|ky|^2\right)
,\tag 3.19$$
and hence, due to $V\in L^1_2({\bold R^+})$ and the boundedness of $\omega(0,y),$ which
follows from (2.28) and (2.35), we
obtain from (3.19) that
$$P_8(k)=o(k^2),\qquad k\to 0 \ {\text{in}} \ {\overline{\bold C^+}}.\tag 3.20$$
As for the integral in (3.14), let us first write it as
$$P_6(k,x)=P_9(-k^*)^\dagger+o(1),\qquad x\to+\infty,\tag 3.21$$
where we have defined
$$P_9(k):=-\int_a^\infty dy\, e^{iky}\,
[\omega(k,y)-\omega(0,y)]\,V(y).\tag 3.22$$
Using (2.25) in (3.22) we get
$$P_9(k)=-ik^2\int_a^\infty dy\, e^{iky}\,V(y)\,\omega_1(y)
-ik^2\int_a^\infty dy\, e^{iky}\,V(y)\,\omega_2(k,y).\tag 3.23$$
Applying (2.31), (2.33), and (2.34) in (3.23), we see that as $k\to 0$ in ${\overline{\bold C^+}}$ we have
$$P_9(k)=k^2
\int_a^\infty dy\,[f(0,x)^\dagger \,f(0,x)-I_n]-
k^2
\int_a^\infty dy\,[f(0,x)-I_n]+o(k^2),$$
or equivalently
$$P_9(-k^*)^\dagger= k^2
\int_a^\infty dy\,[f(0,x)^\dagger \,f(0,x)-I_n]-
k^2
\int_a^\infty dy\,[f(0,x)^\dagger-I_n]+o(k^2).\tag 3.24$$
Finally, using (3.12), (3.15), (3.17), (3.20), (3.21), and (3.24) in (3.5), and
after using in (3.12) the expansion
$$\aligned ik\,e^{ika}f(0,a)^\dagger-e^{ika}f'(0,a)^\dagger=
&(ik-k^2 a)\,f(0,a)^\dagger-\left(1+ika-\displaystyle\frac {k^2 a^2}{2} \right)
f'(0,a)^\dagger
\\
&+
o(k^2),\qquad k\to 0 \text{ in }{\overline{\bold C^+}},\endaligned$$
we obtain (3.3). \qed

One important consequence of (3.3) is given in the following theorem.

\noindent {\bf Theorem 3.2} {\it Assume that the potential $V$ satisfies (1.2)
and belongs to $L^1_2({\bold R}^+).$ Let $f(k,x)$ be the Jost solution to (1.1) appearing
in (2.2). At any fixed $x\in{\bold R^+}$ for which the matrix $f(0,x)$ is invertible,
the matrix $f'(k,x)\,f(k,x)^{-1}$ is twice differentiable
at $k=0$ and we have}
$$\aligned
f'(k,x)\,f(k,x)^{-1}=&f'(0,x)\,f(0,x)^{-1}+ik\,[
f(0,x)^{-1}]^\dagger \,f(0,x)^{-1}\\
&+
k^2\, [f(0,x)^{-1}]^\dagger\, q_1(x)\,
f(0,x)^{-1}+
o(k^2),\qquad k\to 0 \text{ in
}{\overline{\bold C^+}},\endaligned\tag 3.25$$
{\it where we have defined}
$$q_1(x):=-x I_n-i\, f(0,x)^{-1}\,\dot f(0,x)
+\displaystyle\int_x^\infty dy\,[f(0,y)^\dagger f(0,y)-I_n].\tag 3.26$$
{\it At any $x\in{\bold R^+}$ for which the matrix $f'(0,x)$ is invertible,
the matrix $f(k,x)\,f'(k,x)^{-1}$ is twice differentiable
at $k=0$ and we have}
$$\aligned
f(k,x)\,f'(k,x)^{-1}=&f(0,x)\,f'(0,x)^{-1}-ik[
f'(0,x)^{-1}]^\dagger f'(0,x)^{-1}\\
&+k^2\,
[f'(0,x)^{-1}]^\dagger \,q_2(x)\,f'(0,x)^{-1}+
o(k^2),\qquad k\to 0 \text{ in }{\overline{\bold C^+}},\endaligned \tag 3.27$$
{\it where we have defined}
$$q_2(x):=x I_n+i\, f'(0,x)^{-1}\,\dot f'(0,x)
-\displaystyle\int_x^\infty dy\,[f(0,y)^\dagger f(0,y)-I_n].$$

\noindent PROOF:
By Proposition~2.1(a) the matrix-valued function
$f(k,a)$ is continuous in $k\in{\overline{\bold C^+}}$ and hence
the determinant of $f(k,a)$ is also continuous in $k\in{\overline{\bold C^+}}.$
Thus, if $f(0,a)^{-1}$ exists
we must have $f(k,a)$ invertible
in a neighborhood of $k=0$ in ${\overline{\bold C^+}}.$ Then, from (3.2) we get
$$f'(k,a)\,f(k,a)^{-1}=[f(0,a)^\dagger]^{-1}f'(0,a)^\dagger+
[f(0,a)^\dagger]^{-1}P(k)\,f(k,a)^{-1}.\tag 3.28$$
Note that (2.11) implies that
$$[f(0,a)^\dagger]^{-1}f'(0,a)^\dagger=f'(0,a)\,f(0,a)^{-1}.\tag 3.29$$
Applying (3.29) in (3.28) and using (2.12) and (3.3),
we obtain (3.25).
If $f'(0,a)^{-1}$ exists, from (3.2) we get
$$f(k,a)\,f'(k,a)^{-1}=
[f'(0,a)^\dagger]^{-1}\,f(0,a)^\dagger-
[f'(0,a)^\dagger]^{-1}P(k)\, f'(k,a)^{-1}
,$$
and proceeding in a similar way, we obtain
(3.27). \qed

The results in (3.25) and (3.27) are remarkable because the
explicit expansions are given up to $o(k^2),$ whereas the expansions for
$f(k,x)$ and $f'(k,x)$ given in (2.7) and (2.8) can only be obtained up to $o(k).$
 Even though we cannot improve
(2.7) or (2.8) because the existence of $\ddot f(0,x)$ or of $\ddot f'(0,x)$
is not assured when $V\in L^1_2({\bold R^+}),$ we are still able to have the expansions
in (3.25) and (3.27) up to $o(k^2).$

In the scalar case, i.e. when $n=1$ in (1.1), the selfadjointness (1.2) forces
the potential $V$ to be a real-valued scalar function, and as a result
$f(0,x)$ and $f'(0,x)$ also become real valued. Then, from (3.25) and (3.27) we get
the respective remarkable expansions as $k\to 0$ in ${\overline{\bold C^+}}$
$$\aligned\displaystyle\frac{f'(k,x)}{f(k,x)}=& \displaystyle\frac{f'(0,x)}{f(0,x)}+
\displaystyle\frac{ik}{f(0,x)^2}\\
&+ k^2\left[-\displaystyle\frac{i\,\dot f(0,x)}{f(0,x)^3}-
\displaystyle\frac{x}{f(0,x)^2}+\displaystyle\frac{\int_x^\infty dy\,[f(0,y)^2-1]}{f(0,x)^2}
\right]+o(k^2),\endaligned$$
$$\aligned\displaystyle\frac{f(k,x)}{f'(k,x)}=& \displaystyle\frac{f(0,x)}{f'(0,x)}-
\displaystyle\frac{ik}{f'(0,x)^2}\\
&+ k^2\left[\displaystyle\frac{i\,\dot f'(0,x)}{f'(0,x)^3}+
\displaystyle\frac{x}{f'(0,x)^2}-\displaystyle\frac{\int_x^\infty dy\,[f(0,y)^2-1]}{f'(0,x)^2}
\right]+o(k^2).\endaligned$$
We recall that, in the scalar case, $f(0,x)$ and $f'(0,x)$
cannot simultaneously be zero at the same $x$-value
because that would imply that a
corresponding initial-value problem for
(2.5) on the interval $(x,+\infty)$ would only have the zero
solution $f(0,x)\equiv 0,$ contradicting (2.6).

\vskip 10 pt
\noindent {\bf 4. THE JOST MATRIX AND THE SCATTERING MATRIX}
\vskip 3 pt

In this section we recall the definition of the Jost matrix and the scattering
matrix for (1.1) with a selfadjoint matrix potential $V$ in
$L^1_2({\bold R}^+)$ and with the selfadjoint boundary condition
(1.3).

The Jost matrix $J(k)$ for $k\in{\overline{\bold C^+}}$ is defined [5,17,21] as
$$J(k):=[f(-k^*,x)^\dagger;\varphi(k,x)],\tag 4.1$$
where
$f(k,x)$ is the Jost solution to (1.1) appearing in (2.2) and
$\varphi(k,x)$ is the $n\times n$ matrix-valued solution to (1.1) appearing in (2.45).
Since the Wronskian in (4.1) is independent of $x,$
its evaluation at $x=0$ and (2.45) yield
$$J(k)=
f(-k^*,0)^\dagger B-f'(-k^*,0)^\dagger A,\tag 4.2$$
where $A$ and $B$ are the matrices appearing in (1.3) and
(1.4).
Note that $J(k)$ is not uniquely determined by the
potential $V$ and the selfadjoint boundary condition
(1.3). This is because $J(k)$ is unique up to
[5,17] a right multiplication by a constant
invertible matrix, i.e. by a constant matrix
not depending on $k.$
It is known [5,17] that
$J(k)$ is invertible for $k\in{\bold R}\setminus\{0\}$ even under the weaker assumption
$V\in L^1_1({\bold R}^+).$

The scattering matrix $S(k)$ is defined as [5,16-18]
$$S(k):=-J(-k)\,J(k)^{-1},\qquad k\in{\bold R}\setminus\{0\},\tag 4.3$$
and it is uniquely determined by the boundary
condition and the potential $V.$
Even though $J(k)$ is uniquely defined only
up to a right multiplication by a constant invertible
matrix, the
unique determination of $S(k)$ is assured because
$S(k)$ remains unchanged when $J(k)$ is multiplied
on the right by a constant invertible matrix.
We have already established [5] the continuity of $S(k)$ at $k=0$ when the
potential satisfies (1.2) and $V\in L^1_1({\bold R}^+).$
One of our primary goals in the current paper is to establish
the differentiability of $S(k)$ at $k=0$ under the
further assumption $V\in L^1_2({\bold R}^+).$

\vskip 10 pt
\noindent {\bf 5. SMALL-$k$ BEHAVIOR IN THE GENERIC CASE}
\vskip 3 pt

When the potential in (1.1) is selfadjoint and
belongs to $L^1_2({\bold R}^+),$ we are interested in analyzing the behavior of
$J(k)$ and $J(k)^{-1}$ in a neighborhood of $k=0$ in ${\overline{\bold C^+}}$ and the
behavior of $S(k)$ in a neighborhood of $k=0$ in ${\bold R}.$ There are two cases to consider. The first is the {\it generic
case} where $J(0)$ is invertible and the other is the
{\it exceptional
case} where $J(0)$ is not invertible.
The analysis in the exceptional case is delicate, and it will be
given in Section~7. In this section we provide the analysis in the
generic case. We prove the differentiability at $k=0$
for $J(k),$ $J(k)^{-1},$ and $S(k),$ and we also provide their explicit
small-$k$ expansions up to $o(k)$ as $k\to 0.$

\noindent {\bf Theorem 5.1} {\it Assume that the potential $V$ is selfadjoint
and belongs to $L^1_2({\bold R}^+),$ and let $f(k,x)$ be the Jost solution appearing
in (2.2) and $J(k)$ be the associated
Jost matrix defined in (4.1).
Then:}

\item {(a)} {\it The Jost matrix is differentiable at $k=0$ in ${\overline{\bold C^+}}$ and we have}
$$J(k)=J(0)+k\,\dot J(0)+o(k),\qquad k\to 0\ \text{in} \ {\overline{\bold C^+}},\tag 5.1$$
{\it where}
$$J(0)=f(0,0)^\dagger B -f'(0,0)^\dagger A,\tag 5.2$$
$$\dot J(0)=\dot f'(0,0)^\dagger A -\dot f(0,0)^\dagger B,\tag 5.3$$
{\it with $A$ and $B$ being the constant matrices appearing in (1.3).}

\item {(b)} {\it If
$J(0)$ is invertible, then the inverse of the
Jost matrix is differentiable at $k=0$ in ${\overline{\bold C^+}}$ and we have}
$$J(k)^{-1}=J(0)^{-1}-k\,J(0)^{-1}\dot J(0)\, J(0)^{-1}+o(k),\qquad k\to 0\ \text{in} \ {\overline{\bold C^+}},\tag 5.4$$
{\it with $J(0)$ and $\dot J(0)$ given as in (5.2) and (5.3), respectively.}

\item {(c)} {\it If
$J(0)$ is invertible, then the scattering matrix $S(k)$ is differentiable at $k=0$ in ${\bold R}$ and we have}
$$S(k)=-I_n+2 k\,\dot J(0)\, J(0)^{-1}+o(k),\qquad k\to 0\ \text{in} \ {\bold R},\tag 5.5$$
{\it with $J(0)$ and $\dot J(0)$ given as in (5.2) and (5.3), respectively.}

\noindent PROOF: Using (2.7) and (2.8) in (4.2) we obtain (5.1). If $J(0)$ is invertible,
we must have $\det[J(0)]\ne 0.$ Because of the continuity [5]
of $J(k)$ in ${\overline{\bold C^+}},$ the matrix $J(k)^{-1}$ exists in a neighborhood of $k=0$ in
${\overline{\bold C^+}}$ if it exists at $k=0.$ We obtain (5.4) from (5.1) by using
$$J(k)^{-1}=\left[I_n+k\,J(0)^{-1}\dot J(0)+o(k)
\right]^{-1}J(0)^{-1},\qquad k\to 0\ \text{in} \ {\overline{\bold C^+}}.$$
Using (5.1) and (5.4) in (4.3) we establish (5.5). \qed

Let us emphasize that Theorem~5.1(a) holds both in the generic and exceptional cases
because the invertibility of $J(0)$ is not needed there.
We also note that the coefficients in the expansions of (5.1), (5.4), and (5.5) can be constructed from knowledge of the potential $V$ and the constant matrices $A$ and $B.$ This is because $f(0,x)$ and $\dot f(0,x)$ can be obtained by solving (2.5) with the respective
asymptotic conditions (2.14) and (2.18), and hence we directly obtain the constant matrices
$f(0,0)$ and $\dot f(0,0),$ and through differentiation, the
constant matrices $f'(0,0)$ and $\dot f'(0,0).$ Thus, as seen from (5.2) and
(5.3), we have
all the ingredients to determine
the coefficients
in the expansions of (5.1), (5.4), and (5.5).

\vskip 10 pt
\noindent {\bf 6. INVERTIBILITY OF A RELATED MATRIX}
\vskip 3 pt

In this section we continue to analyze the small-$k$ behavior of
the Jost matrix $J(k)$ defined in (4.1). Let us recall that
the $n\times n$ matrix
$J(0)$ is invertible in the generic case but
not invertible in the exceptional case. We show in Theorem~6.1 that the related $2n\times 2n$ matrix $\Cal J$ defined in (6.3)
is always invertible whether we are in the generic case or in the exceptional case.

Let us use $K(k)$ to denote the Jost matrix for (1.1) when
the constant matrices $A$ and $B$ appearing in (1.3) are replaced with
$-B$ and $A,$ respectively. With the help of (5.2) and (5.3), we have
$$K(0)=f(0,0)^\dagger A+f'(0,0)^\dagger B,\tag 6.1$$
$$\dot K(0)=-\dot f(0,0)^\dagger A-\dot f'(0,0)^\dagger B.\tag 6.2$$
Let us define the $2n\times 2n$ matrix $\Cal J$ as
$$\Cal J:=\bmatrix J(0)& K(0)\\
\noalign{\medskip}  \dot J(0) &\dot K(0)\endbmatrix .\tag 6.3$$

The following result shows that $\Cal J$ is always invertible even when $J(0)$ is
not invertible.

\noindent {\bf Theorem 6.1} {\it Assume that the potential
$V$ is selfadjoint and belongs to
$L^1_2({\bold R^+}).$ Then, the matrix $\Cal J$ defined in (6.3) is always invertible
and $\Cal J^{-1}$ is given by}
$$\Cal J^{-1}=\bmatrix i (A^\dagger A+B^\dagger B)^{-1}\dot
K(0)^\dagger& i (A^\dagger A+B^\dagger B)^{-1}
K(0)^\dagger \\
\noalign{\medskip}  -i (A^\dagger A+B^\dagger B)^{-1}
\dot J(0)^\dagger &-i (A^\dagger A+B^\dagger B)^{-1} J(0)^\dagger\endbmatrix .\tag 6.4$$

\noindent PROOF: With the help of (5.2), (5.3), (6.1), and (6.2), we see that
$$\Cal J=
\bmatrix I_n&0\\
\noalign{\medskip}
0& -I_n\endbmatrix\bmatrix f(0,0) & \dot f(0,0)
\\
\noalign{\medskip}
 f'(0,0)& \dot f'(0,0)\endbmatrix ^\dagger
\bmatrix I_n&0\\
\noalign{\medskip}
0& -I_n\endbmatrix
\bmatrix B& A\\
\noalign{\medskip}
A&-B\endbmatrix .\tag 6.5$$
Each matrix on the right-hand side in (6.5) is invertible. The
invertibility of the second matrix on the right-hand side in (6.5)
is assured by (2.19).
 From (1.15) and (1.16) of [5] we have
$$\bmatrix B& A\\
\noalign{\medskip}
A&-B\endbmatrix ^{-1}=
\bmatrix (A^\dagger A+B^\dagger B)^{-1}& 0\\
\noalign{\medskip}
0& (A^\dagger A+B^\dagger B)^{-1}\endbmatrix
\bmatrix B& A\\
\noalign{\medskip}
A&-B\endbmatrix ^\dagger,\tag 6.6$$
where we recall that $A^\dagger A+B^\dagger B$ is positive and hence
invertible. Thus, with the help of (2.19) and (6.6), from (6.5) we get
$$\Cal J^{-1}=-i \bmatrix (A^\dagger A+B^\dagger B)^{-1}& 0\\
\noalign{\medskip}
0& (A^\dagger A+B^\dagger B)^{-1}\endbmatrix
\bmatrix B^\dagger& A^\dagger\\
\noalign{\medskip}
A^\dagger&-B^\dagger\endbmatrix
\bmatrix \dot f'(0,0) &  -f'(0,0)
\\
\noalign{\medskip}
 \dot f(0,0) & -f(0,0)\endbmatrix ,$$
which, by using (5.2), (5.3), (6.1), and (6.2), can also be written as (6.4). \qed

\vskip 10 pt
\noindent {\bf 7. SMALL-$k$ ANALYSIS
OF $J(k)^{-1}$ AND $S(k)$ IN THE EXCEPTIONAL CASE}
\vskip 3 pt

In the exceptional case $J(0)$ is not invertible, and hence we cannot use
(5.4) to express the behavior of $J(k)^{-1}$ as $k\to 0$ in ${\overline{\bold C^+}}$ and
we cannot use (5.5) to express the behavior of $S(k)$ as $k\to 0$ in ${\bold R}.$
Our goal in this section is to obtain such behaviors in the exceptional case when
the potential $V$ in (1.1) satisfies (1.2) and belongs to $V\in L^1_2({\bold R}^+).$
We analyzed such behaviors in [5] under the weaker assumption
$V\in L^1_1({\bold R}^+)$ instead of $V\in L^1_2({\bold R}^+)$ and proved that
$J(k)^{-1}$ has a simple pole at $k=0$ but $S(k)$ is continuous at $k=0.$
In the current paper, under the stronger assumption $V\in L^1_2({\bold R}^+),$
we sharpen the small-$k$ estimates and also prove that
$S(k)$ is differentiable at $k=0.$
Our main results are given in Theorem~7.6, where we prove that
$J(k),$ $k\,J(k)^{-1},$ and $S(k)$ are differentiable at $k=0,$
and in their expansions
as $k\to 0$ we provide the explicit coefficients up to $o(k).$

The following result is already known and the proof is given in Proposition~5.4 of [5].

\noindent {\bf Proposition 7.1} {\it Assume that $V$ in (1.1)
is selfadjoint
and belongs to $L^1_1({\bold R}^+).$ Let $f(k,x),$ $\omega(k,x),$ and $\varphi(k,x)$
be the solutions to (1.1) appearing in (2.2), (2.21), and (2.45), respectively.
Then, the Jost matrix defined in (4.1) can be written as}
$$J(k)=T_1(k)+T_2(k),\qquad k\in{\overline{\bold C^+}},\tag 7.1$$
{\it where we have defined}
$$T_1(k):=-P(-k^*)^\dagger\, f(0,a)^{-1}\,\varphi(k,a),\tag 7.2$$
$$T_2(k):=f(-k^*,a)^\dagger\, [f(0,a)^{-1}]^\dagger\,
[\omega(-k^*,x)^\dagger;\varphi(k,x)],
\tag 7.3$$
{\it with $P(k)$ being the matrix defined in
(3.1) and $a$ being the nonnegative constant
appearing in (2.21).}

The small-$k$ behavior of the
Wronskian $[\omega(-k^*,x)^\dagger;\varphi(k,x)]$
appearing in (7.3) is analyzed in the next proposition
when $V\in L^1_2({\bold R}^+).$

\noindent {\bf Proposition 7.2} {\it Assume that $V$ in (1.1)
is selfadjoint
and belongs to $L^1_2({\bold R}^+).$ Then, the Wronskian appearing in
(7.3) has the small-$k$ asymptotics}
$$[\omega(-k^*,x)^\dagger;\varphi(k,x)]=J(0)+ik^2[\omega'_1(0)^\dagger A-
\omega_1(0)^\dagger B]+o(k^2),
\qquad k\to 0 \text{ in } {\overline{\bold C^+}},\tag 7.4$$
{\it where $A$ and $B$ are the matrices
appearing in (1.3), $J(k)$ is the Jost matrix defined in (4.1), and
$\omega_1(x)$ is the matrix defined in (2.26).}

\noindent PROOF: Since the value of our
Wronskian is independent of
$x,$ we can evaluate its value at $x=0.$
By writing
$$[\omega(-k^*,x)^\dagger;\varphi(k,x)]=[\omega(0,x)^\dagger;\varphi(k,x)]+
[\omega(-k^*,x)^\dagger-\omega(0,x)^\dagger;\varphi(k,x)]
,\tag 7.5$$
 from (2.21), (2.28), (2.45), and (4.2)
we see that the first Wronskian on the right-hand side in (7.5), when
$x=0,$  yields
$$[\omega(0,x)^\dagger;\varphi(k,x)]\big|_{x=0}=J(0).\tag 7.6$$
Next, we evaluate at $x=0$ the value of the
second Wronskian on the right-hand side in (7.5). From (2.27) we see that
$$\omega_2(0,x)=0,\quad \omega'_2(0,x)=0,\qquad x\in{\bold R^+},$$
and
hence (2.25) yields
$$\omega(k,x)-\omega(0,x)=ik^2 \omega_1(x)+o(k^2),
\qquad k\to 0 \text{ in } {\overline{\bold C^+}}.\tag 7.7$$
Using (2.46) and (7.7) with $x=0$ in the second Wronskian
on the right-hand side of (7.5), we get
$$[\omega(-k^*,x)^\dagger-\omega(0,x)^\dagger;\varphi(k,x)]_{x=0}=
ik^2[\omega'_1(0)^\dagger A-
\omega_1(0)^\dagger B]+o(k^2),
\qquad k\to 0 \text{ in } {\overline{\bold C^+}},\tag 7.8$$
and hence (7.6) and (7.8) yield (7.4). \qed

In order to analyze the small-$k$ behavior of $J(k),$ from (7.1), (7.3), and (7.5) we see that it is convenient to analyze the small-$k$ behavior of the related matrix $F(k)$ defined as
$$F(k):=f(0,a)^\dagger [f(-k^*,a)^\dagger]^{-1}
J(k).\tag 7.9$$

\noindent {\bf Proposition 7.3} {\it Assume that $V$ in (1.1)
is selfadjoint
and belongs to $L^1_2({\bold R}^+).$ Then, the $n\times n$ matrix $F(k)$
defined in (7.9) has the small-$k$ asymptotics}
$$F(k)=J(0)-ik\Cal R+k^2 F_2+o(k^2),\qquad k\to 0 \text{ in } {\overline{\bold C^+}},\tag 7.10$$
{\it where we have defined}
$$\Cal R:=  f(0,a)^{-1}\varphi(0,a),\tag 7.11$$
$$F_2:=i \, [\omega'_1(0)^\dagger A-
\omega_1(0)^\dagger B]-q_1(a)^\dagger \,\Cal R,\tag 7.12$$
{\it with $f(k,x)$ and $\varphi(k,x)$ being the
solutions to (1.1) and satisfying (2.2) and (2.45),
respectively, and where $\omega_1(x)$ and $q_1(x)$ are the quantities defined in
(2.26) and (3.26), respectively,
and $a$ is the nonnegative constant appearing in (2.21).}

\noindent PROOF: From (7.1)-(7.3) we see that
$$F(k)=-f(0,a)^\dagger\,[ f(-k^*,a)^\dagger]^{-1} \,P(-k^*)^\dagger \, f(0,a)^{-1}\varphi(k,a)+[\omega(-k^*,x)^\dagger;\varphi(k,x)].\tag 7.13$$
Using (2.12), (2.46), (3.3), and (7.4) in (7.13), we get the expansion in (7.10). \qed

Since $J(0)$ is not invertible in the exceptional case, in order to analyze
the small-$k$ behavior of $J(k)^{-1},$ we will
transform (7.10) into the representation where the matrix $J(0)$ is in
its Jordan canonical form with the zero eigenvalues appearing in the upper blocks.
We refer the reader to Section~VI of [5], and in particular to (6.7) of
[5], where such a similarity transformation is accomplished by the invertible constant matrix $\Cal S$ constructed with the help of
the eigenvectors and generalized eigenvectors associated with
the zero eigenvalue of the matrices $J(0)$ and
$J(0)^\dagger.$

As in (6.12) of [5], let us introduce
$$\Cal Z(k):=
\bmatrix \Cal A(k)&\Cal B(k)\\
\noalign{\medskip}
\Cal C(k)&\Cal D(k)\endbmatrix :
=P_2\,\Cal S^{-1}F(k)\,\Cal S\,P_1,
\tag 7.14$$
where $\Cal A(k)$ has size $\mu\times \mu$ and
$\Cal D(k)$ has size $(n-\mu)\times (n-\mu),$
and $P_1$ and $P_2$
are the permutation
matrices defined in (6.11) of [5].
Here, $\mu$ is the
geometric multiplicity of the zero eigenvalue of $J(0).$
The matrices $P_1$ and $P_2$ help to
move the elements $1$ appearing in the Jordan canonical form
of $J(0)$ from the superdiagonal to the diagonal
so that
the Jordan block form of $J(0)$ is transformed into the form given in (7.17),
where $P_1$ permutes the rows and $P_2$ permutes the columns of
the Jordan canonical form
of $J(0).$

\noindent {\bf Proposition 7.4} {\it Assume that $V$ in (1.1)
is selfadjoint
and belongs to $L^1_2({\bold R}^+).$ Then:}

\item{(a)} {\it The small-$k$ asymptotics as $k\to 0$ in ${\overline{\bold C^+}}$
of the $n\times n$ matrix $\Cal Z(k)$
defined in (7.14) is given by}
$$\Cal Z(k)=P_2\Cal S^{-1} J(0)\, \Cal S P_1-ik
P_2\Cal S^{-1} \Cal R \Cal S P_1+k^2
P_2\Cal S^{-1} F_2 \Cal S P_1+o(k^2),\tag 7.15$$
{\it where $\Cal R$ and $F_2$ are the constant $n\times n$ matrices defined
in (7.11) and (7.12), respectively.}

\item{(b)} {\it In the exceptional case, i.e. when
$J(0)$ is not invertible, the expansion in (7.15) is equivalent to the expansion
as $k\to 0$ in ${\overline{\bold C^+}}$}
$$\bmatrix \Cal A(k)&\Cal B(k)\\
\noalign{\medskip}
\Cal C(k)&\Cal D(k)\endbmatrix =\bmatrix k\,\Cal A_1+k^2\,\Cal A_2+o(k^2)&
k\,\Cal B_1+k^2\,\Cal B_2+o(k^2)
\\
\noalign{\medskip}
k\,\Cal C_1+k^2\,\Cal C_2+o(k^2)&\Cal D_0+
k\,\Cal D_1+k^2\,\Cal D_2+o(k^2)\endbmatrix ,\tag 7.16$$
{\it with $\Cal A_1$ and $\Cal D_0$ being constant invertible
matrices of sizes $\mu\times \mu$ and $(n-\mu)\times (n-\mu),$
respectively, and where $\Cal A_2,$ $\Cal B_1,$ $\Cal B_2,$
$\Cal C_1,$ $\Cal C_2,$ $\Cal D_1,$ and $\Cal D_2$
are some constant matrices.}

\noindent PROOF: We obtain (7.15) directly from (7.10) by multiplying on the left
 and on the right with the constant matrices $P_2\Cal S^{-1}$ and
$\Cal S P_1,$ respectively. We refer the reader to Section~VI of [5] for the expansion in (7.15) without the term containing $k^2$ when the potential satisfies the weaker assumption $V\in L^1_1({\bold R^+}).$ In the expansion in (7.15), the constant term
$P_2\Cal S^{-1} J(0) \Cal S P_1$ has the form given in the displayed formula
before (6.1) of [5], namely
$$P_2\Cal S^{-1} J(0) \Cal S P_1=\text{diag}\{0_\mu,I_{\nu-\mu},
 J_{n_{\mu+1}}(\lambda_{\mu+1})
,\dots
,J_{n_\kappa}(\lambda_{\kappa})\}.\tag 7.17$$
In the block diagonal matrix given on the right-hand side of
(7.17) the zero block matrix $0_\mu$ has size $\mu\times\mu,$ where we recall that
$\mu$ is the
geometric multiplicity of the zero eigenvalue of $J(0).$ The identity
matrix $I_{\nu-\mu}$ in (7.17) has size $(\nu-\mu)\times (\nu-\mu),$ where
$\nu$ denotes the algebraic multiplicity of the zero eigenvalue of $J(0).$
The remaining block matrices $J_{n_\alpha}(\lambda_{\alpha})$
for $\alpha=\mu+1,\dots,\kappa$ are the Jordan blocks associated with the
nonzero eigenvalues of $J(0),$ as described in (6.7) of [5].
Let us use $\Cal D_0$ to denote the nonzero block diagonal matrix
of size $(n-\mu)\times (n-\mu)$ in (7.17), i.e.
$$\Cal D_0:=\text{diag}\{I_{\nu-\mu},
 J_{n_{\mu+1}}(\lambda_{\mu+1})
,\dots
,J_{n_\kappa}(\lambda_{\kappa})\}.\tag 7.18$$
Since none of the eigenvalues $\lambda_{\alpha}$
for $\alpha=\mu+1,\dots,\kappa$
appearing in (7.18) are zero,
the matrix $\Cal D_0$ is invertible.
The term proportional to $k$ in (7.16) is already known and has been analyzed in Theorem~6.1 of [5] under the weaker assumption $V\in L^1_1({\bold R^+}).$
As in (6.13) of [5], we have
$$\bmatrix \Cal A_1&\Cal B_1\\
\noalign{\medskip}
\Cal C_1&\Cal D_1\endbmatrix:=-i
P_2\Cal S^{-1} \Cal R \Cal S P_1,\tag 7.19$$
where $\Cal A_1$ is invertible, which is proved in Theorem~6.1 of [5].
For the term proportional to $k^2$ in the expansion of (7.17), let us
introduce the block matrix notation
$$\bmatrix \Cal A_2&\Cal B_2\\
\noalign{\medskip}
\Cal C_2&\Cal D_2\endbmatrix:=P_2\Cal S^{-1} F_2 \Cal S P_1,\tag 7.20$$
where $\Cal A_2$ has size $\mu\times \mu$ and $\Cal D_2$
has size $(n-\mu)\times (n-\mu)$ and hence $\Cal B_2$ and $\Cal C_2$
have sizes $\mu\times (n-\mu)$ and $(n-\mu)\times\mu,$
respectively.
Using (7.14) and (7.17)-(7.20) in (7.15) we
obtain (7.16). \qed

Next, we analyze the small-$k$ asymptotics for the inverse of the matrix $\Cal Z(k)$ defined
in (7.14).

\noindent {\bf Proposition 7.5} {\it Assume that the potential $V$ in (1.1)
is selfadjoint
and belongs to $L^1_2({\bold R}^+),$ and let $\Cal Z(k)$ be the $n\times n$ matrix
defined in (7.14). In the exceptional case, i.e. when
$J(0)$ is not invertible, $\Cal Z(k)^{-1}$ has the small-$k$ asymptotics}
as $k\to 0$ in ${\overline{\bold C^+}}$
$$\Cal Z(k)^{-1}=\bmatrix \displaystyle\frac{1}{k}\,\Cal A_1^{-1}+\Cal Y_1+o(1)&
-\Cal A_1^{-1}\Cal B_1\Cal D_0^{-1}+k \Cal Y_2+o(k)\\
\noalign{\medskip}
-\Cal D_0^{-1}\Cal C_1\Cal A_1^{-1}+k \Cal Y_3+o(k)&
\Cal D_0^{-1}+k \Cal Y_4+O(k^2)\endbmatrix ,
\tag 7.21$$
{\it where we have defined, in terms of the matrices appearing in (7.18)-(7.20),}
$$\Cal Y_1:=\Cal A_1^{-1}(\Cal B_1\Cal D_0^{-1}\Cal C_1-\Cal A_2)\Cal A_1^{-1},$$
$$\Cal Y_2:=\Cal A_1^{-1}[\Cal A_2\Cal A_1^{-1}\Cal B_1-\Cal B_1\Cal D_0^{-1}\Cal C_1\Cal A_1^{-1}\Cal B_1-\Cal B_2-\Cal B_1\Cal D_0^{-1}\Cal D_1]\Cal D_0^{-1}
,$$
$$\Cal Y_3:=\Cal D_0^{-1}[\Cal D_1\Cal D_0^{-1}\Cal C_1-\Cal C_2+\Cal C_1\Cal A_1^{-1}\Cal A_2-\Cal C_1\Cal A_1^{-1}\Cal B_1\Cal D_0^{-1}\Cal C_1
]\Cal A_1^{-1},$$
$$\Cal Y_4:=\Cal D_0^{-1}(\Cal C_1 \Cal A_1^{-1}\Cal B_1-\Cal D_1)\Cal D_0^{-1}.$$

\noindent PROOF: We can readily evaluate the inverse of
$\Cal Z(k)$ appearing in (7.14) by using
(6.16) of [5], namely
$$\bmatrix \Cal A&\Cal B\\
\noalign{\medskip}
\Cal C&\Cal D\endbmatrix ^{-1}=\bmatrix
(\Cal A-\Cal B\Cal D^{-1}\Cal C)^{-1}&-(\Cal A-\Cal B\Cal D^{-1}\Cal C)^{-1}
\Cal B\Cal D^{-1}
\\
\noalign{\medskip}
-\Cal D^{-1}\Cal C(\Cal A-\Cal B\Cal D^{-1}\Cal C)^{-1}&
\Cal D^{-1}\Cal C(\Cal A-\Cal B\Cal D^{-1}\Cal C)^{-1}\Cal B\Cal D^{-1}+\Cal D^{-1}\endbmatrix ,\tag 7.22$$
as well as the expansions as $k\to 0$ in ${\overline{\bold C^+}}$
$$[k\,\Cal A_1+k^2\,\Cal A_2+o(k^2)]^{-1}=
\displaystyle\frac{1}{k}\,\Cal A_1^{-1}-\Cal A_1^{-1} \Cal A_2 \Cal A_1^{-1}+o(1),
\tag 7.23$$
$$\aligned
\left[\Cal D_0+
k\,\Cal D_1+k^2\,\Cal D_2+o(k^2)\right]^{-1}=&\Cal D_0^{-1}-k\,\Cal D_0^{-1} \Cal D_1 \Cal D_0^{-1}\\
&+k^2[\Cal D_0^{-1} \Cal D_1 \Cal D_0^{-1}\Cal D_1 \Cal D_0^{-1}-\Cal D_0^{-1} \Cal D_2 \Cal D_0^{-1}
]+o(k^2),\endaligned
\tag 7.24$$
where we recall that the invertibility of $\Cal A_1$ and $\Cal D_0$ is already known [5].
Using (7.16), (7.24), and an analog of (7.23) on
the right-hand side of (7.22), we obtain (7.21). \qed

We are now ready to evaluate the small-$k$ asymptotics of
the Jost matrix $J(k)$ defined in (4.1), its inverse $J(k)^{-1},$ and
the scattering matrix $S(k)$ defined in (4.3). From (7.9) and
(7.14) we see that
the Jost matrix $J(k)$
is given by
$$J(k)=f(-k^*,a)^\dagger\,[f(0,a)^\dagger]^{-1}
\Cal S P_2^{-1} \Cal Z(k)\,P_1^{-1} \Cal S^{-1},\tag 7.25$$
where the small-$k$ asymptotics in ${\overline{\bold C^+}}$ will be evaluated with the help
of (7.16). On the other hand, from  (7.25)
we get
$$J(k)^{-1}=
\Cal S \,P_1\,\Cal Z(k)^{-1}P_2\,  \Cal S^{-1}f(0,a)^\dagger
\,[f(-k^*,a)^\dagger]^{-1},\tag 7.26$$
where the small-$k$ asymptotics in ${\overline{\bold C^+}}$ will be evaluated with the help
of (7.21).
Using (7.25) and (7.26) in (4.3) we obtain for $k\in{\bold R}\setminus\{0\}$
$$S(k)=-f(k,a)^\dagger\,[f(0,a)^\dagger]^{-1}
\Cal S P_2^{-1} \Cal Z(-k)\,\Cal Z(k)^{-1}P_2 \, \Cal S^{-1}f(0,a)^\dagger
\,[f(-k,a)^\dagger]^{-1},\tag 7.27$$
where the small-$k$ asymptotics will be evaluated with the help
of (7.16) and (7.21).
In evaluating the small-$k$ asymptotics of
$J(k),$ $J(k)^{-1},$ and $S(k),$
we will also use some expansions related to (2.7) and (2.12), namely
$$f(-k^*,a)^\dagger[f(0,a)^\dagger]^{-1}=I_n-k\,\dot f(0,a)^\dagger[f(0,a)^\dagger]^{-1}+ o(k),\qquad k\to 0 \ \text{in} \ {\overline{\bold C^+}},\tag 7.28$$
$$f(0,a)^\dagger\,[f(-k^*,a)^\dagger]^{-1}=I_n+k\, \dot f(0,a)^\dagger [ f(0,a)^\dagger]^{-1}+o(k),\qquad k\to 0 \ \text{in} \ {\overline{\bold C^+}}.\tag 7.29$$

\noindent {\bf Theorem 7.6} {\it Assume that $V$ in (1.1)
is selfadjoint
and belongs to $L^1_2({\bold R}^+).$ Let
$J(k)$ be the Jost matrix defined in (4.1) and assume that we are in the
exceptional case, i.e.
$J(0)$ is not invertible. Then:}

\item{(a)} {\it The Jost matrix $J(k)$ is differentiable at $k=0$
in ${\overline{\bold C^+}}$ and has the behavior}
$$J(k)=J(0)+k\,\dot J(0)+o(k), \qquad k\to 0 \ \text{in} \ {\overline{\bold C^+}},\tag 7.30$$
{\it where $J(0)$ and $\dot J(0)$ are given,
in terms of the matrices appearing in (7.17)-(7.19), by}
$$J(0)=\Cal S P_2^{-1}
\bmatrix 0& 0\\
\noalign{\medskip}
0&\Cal D_0\endbmatrix
P_1^{-1}  \Cal S^{-1},\tag 7.31$$
$$\dot J(0)=-\dot f(0,a)^\dagger\,[f(0,a)^\dagger]^{-1}J(0)+\Cal S P_2^{-1}
\bmatrix \Cal A_1& \Cal B_1\\
\noalign{\medskip}
\Cal C_1&\Cal D_1\endbmatrix
P_1^{-1}  \Cal S^{-1}.\tag 7.32$$

\item{(b)} {\it The Jost matrix $J(k)$ has a simple pole at $k=0,$ and we
have the asymptotics}
$$J(k)^{-1}=\displaystyle\frac{1}{k}\, \Cal S P_1
\bmatrix  \Cal A_1^{-1}&0\\
\noalign{\medskip}
0&0\endbmatrix P_2 \Cal S^{-1}+\Cal E_1+o(1),
\qquad k\to 0 \ {\text{in}} \ {\overline{\bold C^+}},\tag 7.33$$
{\it where we have defined}
$$\aligned \Cal E_1:=&
\Cal S P_1
\bmatrix  \Cal A_1^{-1}&0\\
\noalign{\medskip}
0&0\endbmatrix P_2 \Cal S^{-1} \dot f(0,a)^\dagger [f(0,a)^\dagger]^{-1}\\
&+
\Cal S P_1\bmatrix
\Cal A_1^{-1}(\Cal B_1\Cal D_0^{-1}\Cal C_1-\Cal A_2)\Cal A_1^{-1}&
-\Cal A_1^{-1}\Cal B_1\Cal D_0^{-1}\\
\noalign{\medskip}
-\Cal D_0^{-1}\Cal C_1\Cal A_1^{-1}&
\Cal D_0^{-1}
\endbmatrix P_2\Cal S^{-1}
.\endaligned\tag 7.34$$

\item{(c)} {\it The scattering matrix $S(k)$ defined in (4.3)
is differentiable at $k=0$ in ${\bold R}$ and satisfies the asymptotics}
$$S(k)=S(0)+k\,\dot S(0)+o(k),
\qquad k\to 0 \ {\text{in}} \ {\bold R},\tag 7.35$$
{\it where we have}
$$S(0)=\Cal S P_2^{-1}
\bmatrix I_\mu& 0\\
\noalign{\medskip}
2\Cal C_1\Cal A_1^{-1}&-I_{n-\mu}\endbmatrix
P_2  \Cal S^{-1},\tag 7.36$$
$$\dot S(0)=\Cal S P_2^{-1} \Cal E_2 P_2
\Cal S^{-1}+S(0)\,
\dot f(0,a)^\dagger [f(0,a)^\dagger]^{-1}
+\dot f(0,a)^\dagger[f(0,a)^\dagger]^{-1}S(0)
,\tag 7.37$$
{\it with the $n\times n$ matrix $\Cal E_2$ defined as}
$$\Cal E_2:=2
\bmatrix  -\Cal A_2 \Cal A_1^{-1}&0\\
\noalign{\medskip}
\Cal E_3&
(\Cal D_1-\Cal C_1\Cal A_1^{-1}\Cal B_1)\Cal D_0^{-1}
\endbmatrix ,\tag 7.38$$
{\it and with the $(n-\mu)\times\mu$ matrix $\Cal E_3$ given by}
$$\Cal E_3:=(\Cal C_1\Cal A_1^{-1}\Cal B_1\Cal D_0^{-1}\Cal C_1-
C_1\Cal A_1^{-1}\Cal A_2-\Cal D_1\Cal D_0^{-1}\Cal C_1
)\Cal A_1^{-1}.\tag 7.39$$

\noindent PROOF: In order to get (7.30), we use (7.14), (7.16), and (7.28) in (7.25).
In a similar way, we obtain (7.33) by using (7.21) and (7.29) in (7.26). Finally,
we obtain (7.35) by using (7.16), (7.21), (7.28), and (7.29) in (7.27). \qed

Let us emphasize that in the expansions (7.30), (7.33), and (7.35) for
$J(k),$ $J(k)^{-1},$ and $S(k),$ respectively, we can construct explicitly the first two terms
 from knowledge of the
matrix potential $V$ and the constant matrices $A$ and $B$ appearing in (1.3).
For this purpose, we can first obtain $f(0,x)$ and $\dot f(0,x)$ by solving
(2.5) with the respective asymptotic conditions (2.14) and (2.18). We then
get $J(0)$ from (5.2). Next, we can construct the constant matrices $P_1,$ $P_2,$ and
$\Cal S$ so that (7.17) is satisfied. Using (3.26),
we can obtain the constant matrix $q_1(a),$ where $a$ is a nonnegative
number for which the matrix $f(0,a)$ is invertible. Next,
with the help of (2.38) and (2.39), we can get the
constant matrices $\omega_1(0)$ and $\omega_1'(0).$
We can construct $\varphi(0,a)$ by solving (2.5) with the conditions in
(2.45) by putting $k=0$ there. Then, via (7.11) and (7.12) we can
get the constant matrices $\Cal R$ and $F_2.$ Next, with
the help of (7.18)-(7.20), we obtain the constant matrices $\Cal A_1,$ $\Cal A_2,$
$\Cal B_1,$ $\Cal C_1,$ $\Cal D_0,$ and $\Cal D_1.$
We then have all the ingredients for the first two terms
in each of the expansions (7.30), (7.33), and (7.35) because, as seen from (7.30)-(7.39), those
coefficients can explicitly be obtained from the constant matrices
$f(0,a),$ $\dot f(0,a),$ $P_1,$ $P_2,$ $\Cal S,$ $\Cal A_1,$ $\Cal A_2,$ $\Cal B_1,$ $\Cal C_1,$ $\Cal D_0,$ and $\Cal D_1.$

\vskip 10 pt
\noindent {\bf 8. AN EXPLICIT EXAMPLE}
\vskip 3 pt

In this section we illustrate our results on the small-$k$ asymptotics of
the Jost matrix, the inverse of the Jost matrix,
 and the scattering matrix by presenting an explicit example.
In the generic case we illustrate and verify the asymptotics in (5.1), (5.4), and
(5.5). In the exceptional case we illustrate and verify
the asymptotics given in (7.30), (7.33), and (7.35).

In our example we use the so-called Kirchhoff boundary conditions
given by
$$\psi_1(0)=\psi_2(0)=\psi_3(0),
\quad \psi'_1(0)+\psi'_2(0)+\psi'_3(0)=0,\tag 8.1$$
where $\psi_j(x)$ denotes the $j$th component of the wavefunction $\psi(x)$
appearing in (1.3). Such boundary conditions are relevant
in many application areas and may correspond,
for example in quantum wires, to the continuity of the wavefunction and the conservation of
the current at a junction.

Assume that the potential is given by
$$V(x)=\bmatrix \displaystyle\frac{32e^{2x}}{(4e^{2x}-1)^2}&0&0\\
\noalign{\medskip}
0&0&0\\
\noalign{\medskip}
0&0&0\endbmatrix +\delta (x-1)\bmatrix 0&0&0\\
\noalign{\medskip}
0&1&1\\
\noalign{\medskip}
0&1 &\gamma \endbmatrix ,$$
where $\delta(x)$ denotes the Dirac delta distribution and $\gamma$ is a real parameter. We will see that
the choice $\gamma=-31/77$ corresponds to $\det[J(0)]=0$ and any other $\gamma$ value corresponds to $\det[J(0)]\ne 0.$
Suppose that the boundary
parameter matrices $A$ and $B$ appearing in (1.3) are given by
$$A=\bmatrix 0&0&1\\
\noalign{\medskip}
0&0&1\\
\noalign{\medskip}
0&0&1\endbmatrix ,\quad B=\bmatrix -1&0&0,\\
\noalign{\medskip}
1&-1&0\\
\noalign{\medskip}
0&1&0\endbmatrix ,\tag 8.2$$
so that we have the Kirchhoff boundary conditions given
in (8.1).
We can evaluate the
Jost solution explicitly and obtain
$$f(k,x)=
\cases \bmatrix f_1& 0 &0\\
\noalign{\medskip}
0& f_2 & f_3
\\
\noalign{\medskip}
0& f_3 & f_4\endbmatrix ,\qquad 0\le x\le 1,\\
\bmatrix f_1& 0 &0\\
\noalign{\medskip}
0& e^{ikx}&0
\\
\noalign{\medskip}
0& 0 & e^{ikx}\endbmatrix ,\qquad x\ge 1,\endcases\tag 8.3$$
where we have defined
$$f_1:=e^{ikx}\left(1+\displaystyle\frac{2i}{(k+i)(4e^{2x}-1)}\right),\quad
f_2:= e^{ikx}\left(1+\displaystyle\frac{i}{2k}\right)-\displaystyle\frac{i}{2k}\,
e^{ik(2-x)},$$
$$f_3:=\displaystyle\frac{i}{2k}\,e^{ikx}-\displaystyle\frac{i}{2k}\,
e^{ik(2-x)},\quad
f_4:=e^{ikx}\left(1+\displaystyle\frac{i\gamma}{2k}\right)-\displaystyle\frac{i\gamma}{2k}\,
e^{ik(2-x)}.$$
Note that $f(k,0)$ and $f'(k,0)$ are given by
$$f(k,0)=\bmatrix 1+\displaystyle\frac{2i}{3(k+i)}&0&0\\
\noalign{\medskip}
0&1+\displaystyle\frac{i}{2k}\left(1-e^{2ik}\right)&\displaystyle\frac{i}{2k}\left(1-e^{2ik}\right)\\
\noalign{\medskip}
0&\displaystyle\frac{i}{2k}\left(1-e^{2ik}\right)&1+\displaystyle\frac{i\gamma}{2k}\left(1-e^{2ik}\right)\endbmatrix ,\tag 8.4$$
$$f'(k,0)=\bmatrix ik-\displaystyle\frac{2(3k+8i)}{9(k+i)}&0&0\\
\noalign{\medskip}
0&ik-\displaystyle\frac{1}{2}\left(1+e^{2ik}\right)&-e^{ik}\cos k\\
\noalign{\medskip}
0&-e^{ik}\cos k& ik-\displaystyle\frac{\gamma}{2}\left(1+e^{2ik}\right)\endbmatrix .\tag 8.5$$
Using (8.3)-(8.5) in (4.2) we get the Jost matrix as
$$J(k)=\bmatrix -1-\displaystyle\frac{2i}{3(k+i)}&0& -ik+\displaystyle\frac{2}{3}+\displaystyle\frac{10i}{9(k+i)}\\
\noalign{\medskip} 1+\displaystyle\frac{i}{2k}\left(1-e^{2ik}\right)&-1&1-ik+ e^{2ik}\\
\noalign{\medskip} \displaystyle\frac{1}{k}\,e^{ik}\sin k&1  +\displaystyle\frac{\gamma-1}{k}\,e^{ik}\sin k &-ik+\displaystyle\frac{\gamma+1}{2}\left(1+e^{2ik}\right)  \endbmatrix .$$
The determinant of $J(k)$ has the small-$k$ asymptotics
$$\det[J(k)]=\displaystyle\frac{31+77\gamma}{9}+\displaystyle\frac{(128\gamma-39)ik}{9}+
\left(\displaystyle\frac{131}{27}-\displaystyle\frac{193\gamma}{9}\right)k^2+O(k^3),\qquad k\to 0 \text{ in } \bold C.
$$
Thus, $\gamma=-31/77$ corresponds to the exceptional case, i.e. $\det[J(0)]=0$, and any other value of $\gamma$ yields a generic
case, i.e. $\det[J(0)]\ne 0.$
The small-$k$ limit of $J(k)$ is given by
$$J(k)=J(0)+k\,\dot J(0)+O(k^2),\qquad k\to 0 \text{ in } {\bold C},\tag 8.6$$
with
$$J(0)=\bmatrix -\displaystyle\frac{5}{3}&0&\displaystyle\frac{16}{9}\\
\noalign{\medskip} 2&-1&2\\
\noalign{\medskip}
1&\gamma &1+\gamma \endbmatrix ,\quad
\dot J(0)=\bmatrix -\displaystyle\frac{2i}{3}&0&\displaystyle\frac{i}{9}\\
\noalign{\medskip} i&0&i\\
\noalign{\medskip}
i&i(\gamma -1)&i\gamma \endbmatrix .\tag 8.7$$
For the choice $\gamma =-31/77$ we get the asymptotics
$$J(k)^{-1}=\displaystyle\frac{i}{k}\,M_1+M_2+O(k),\qquad k\to 0 \text{ in } {\bold C},\tag 8.8$$
where we have defined
$$M_1:=\bmatrix \displaystyle\frac{144}{6971}&-\displaystyle\frac{496}{6971}&\displaystyle\frac{1232}{6971}\\
\noalign{\medskip}
\displaystyle\frac{558}{6971}&-\displaystyle\frac{1922}{6971}&\displaystyle\frac{4774}{6971}\\
\noalign{\medskip}
\displaystyle\frac{135}{6971}&-\displaystyle\frac{465}{6971}&\displaystyle\frac{1155}{6971}\endbmatrix ,\quad
M_2:=
\bmatrix -\displaystyle\frac{16095714}{48594841}&
\displaystyle\frac{22880111}{145784523}&\displaystyle\frac{32930051}{145784523}\\
\noalign{\medskip}
-\displaystyle\frac{10281837}{48594841}&-\displaystyle\frac{30462632}{145784523}&\displaystyle\frac{61380319}{145784523}\\
\noalign{\medskip}
\displaystyle\frac{11927250}{48594841}&\displaystyle\frac{8244046}{48594841}
&\displaystyle\frac{7573258}{48594841}\endbmatrix ,$$
and similarly, for the scattering matrix we get
$$S(k)=S(0)+k\,\dot S(0)+O(k^2),\qquad k\to 0 \text{ in } {\bold C},\tag 8.9$$
with
$$S(0)=\bmatrix -\displaystyle\frac{6809}{6971}&-\displaystyle\frac{558}{6971}&\displaystyle\frac{1386}{6971}\\
\noalign{\medskip}
-\displaystyle\frac{558}{6971}&-\displaystyle\frac{5049}{6971}&-\displaystyle\frac{4774}{6971}\\
\noalign{\medskip}
\displaystyle\frac{1386}{6971}&-\displaystyle\frac{4774}{6971}&\displaystyle\frac{4887}{6971}\endbmatrix ,
\tag 8.10$$
$$
\dot S(0)=\bmatrix \displaystyle\frac{24111452 i}{48594841}&
-\displaystyle\frac{8336928 i}{48594841}&-\displaystyle\frac{12952632 i}{48594841}\\
\noalign{\medskip}
-\displaystyle\frac{8336928 i}{48594841}&\displaystyle\frac{95224498 i}{145784523}&\displaystyle\frac{111299650 i}{145784523}\\
\noalign{\medskip}
-\displaystyle\frac{12952632 i}{48594841}&\displaystyle\frac{111299650 i}{145784523}
&-\displaystyle\frac{124617494 i}{145784523}\endbmatrix .\tag 8.11$$
In the generic case, we have (8.9) with the relevant matrices
specified as
$$S(0)=-I_3,\quad
\dot S(0)=\bmatrix \displaystyle\frac{2i(20\gamma +7)}{77\gamma +31}&-\displaystyle\frac{18i\gamma }{77\gamma +31}&-\displaystyle\frac{18i}{77\gamma +31}\\
\noalign{\medskip}
-\displaystyle\frac{18i\gamma }{77\gamma +31}&\displaystyle\frac{62i\gamma }{77\gamma +31}&\displaystyle\frac{62i}{77\gamma +31}\\
\noalign{\medskip}
-\displaystyle\frac{18i}{77\gamma +31}&\displaystyle\frac{62i}{77\gamma +31}&\displaystyle\frac{2i(77\gamma -46)}{77\gamma +31}\endbmatrix ,\tag 8.12$$
where we recall that $I_3$ denotes the $3\times 3$ identity matrix.

Let us now verify our results on the small-$k$ limits. In the generic case, i.e. when $\gamma \ne -31/77,$
we will obtain the small-$k$ limits as described in Theorem~5.1. From (8.4) and (8.5) we get
$$f(0,0)=\bmatrix \displaystyle\frac{5}{3}&0&0\\
\noalign{\medskip}
0&2&1\\
\noalign{\medskip}
0&1&\gamma +1\endbmatrix ,\quad f'(0,0)=\bmatrix -\displaystyle\frac{16}{9}&0&0\\
\noalign{\medskip}
0&-1&-1\\
\noalign{\medskip}
0&-1&-\gamma \endbmatrix ,\tag 8.13$$
$$\dot f(0,0)= \bmatrix \displaystyle\frac{2i}{3}&0&0\\
\noalign{\medskip}
0&i&i\\
\noalign{\medskip}
0&i&i\gamma \endbmatrix ,\quad
\dot f'(0,0)=\bmatrix -\displaystyle\frac{i}{9}&0&0\\
\noalign{\medskip}
0&0&-i\\
\noalign{\medskip}
0&-i&i(1-\gamma )\endbmatrix.\tag 8.14$$
Using (8.2), (8.13), and (8.14) in (5.2) and (5.3) we obtain (8.6) with
$$J(0)=\bmatrix -\displaystyle\frac{5}{3}&0&\displaystyle\frac{16}{9}\\
\noalign{\medskip} 2&-1&2\\
\noalign{\medskip}
1&\gamma &1+\gamma \endbmatrix ,\quad
\dot J(0)=\bmatrix -\displaystyle\frac{2i}{3}&0&\displaystyle\frac{i}{9}\\
\noalign{\medskip} i&0&i\\
\noalign{\medskip}
i&i(\gamma -1)&i\gamma \endbmatrix ,\tag 8.15$$
agreeing with
the quantities given in (8.7). Similarly, using (8.15)
in (5.5) we obtain (8.9)
with $S(0)$ and $\dot S(0)$ agreeing with the quantities given in (8.10) and (8.11), respectively.
Thus, we have verified the results of Theorem~5.1 in the generic case for our example.

Let us now verify, in the exceptional case i.e. when $\gamma =-31/77,$ the small-$k$ asymptotics presented in Section~7. From (8.4) and (8.5) with $\gamma =-31/77,$ we obtain
$$f(0,0)=\bmatrix \displaystyle\frac{5}{3}&0&0\\
\noalign{\medskip}
0&2&1\\
\noalign{\medskip}
0&1&\displaystyle\frac{46}{77}\endbmatrix ,\quad f'(0,0)=\bmatrix -\displaystyle\frac{16}{9}&0&0\\
\noalign{\medskip}
0&-1&-1\\
\noalign{\medskip}
0&-1&\displaystyle\frac{31}{77}\endbmatrix ,\tag 8.16$$
$$\dot f(0,0)= \bmatrix \displaystyle\frac{2i}{3}&0&0\\
\noalign{\medskip}
0&i&i\\
\noalign{\medskip}
0&i&-\displaystyle\frac{31i}{77}\endbmatrix ,\quad
\dot f'(0,0)=\bmatrix -\displaystyle\frac{i}{9}&0&0\\
\noalign{\medskip}
0&0&-i\\
\noalign{\medskip}
0&-i&\displaystyle\frac{108i}{77}\endbmatrix.\tag 8.17$$
Using (8.2), (8.4), and (8.5) in (5.2) and (5.3), we obtain
(8.6) with
$$J(0)=\bmatrix -\displaystyle\frac{5}{3}&0&\displaystyle\frac{16}{9}\\
\noalign{\medskip} 2&-1&2\\
\noalign{\medskip}
1&-\displaystyle\frac{31}{77}&\displaystyle\frac{46}{77}\endbmatrix ,\quad
\dot J(0)=\bmatrix -\displaystyle\frac{2i}{3}&0&\displaystyle\frac{i}{9}\\
\noalign{\medskip} i&0&i\\
\noalign{\medskip}
i&-\displaystyle\frac{108i}{77}&-\displaystyle\frac{31i}{77}\endbmatrix ,\tag 8.18$$
agreeing with the values given in (8.7) when $\gamma =-31/77.$
The eigenvalues of $J(0)$ are obtained from (8.15) and are given by
$$\lambda_1=0,\quad \lambda_2=\displaystyle\frac{-239+2\sqrt{26273}}{231},\quad
\lambda_3=\displaystyle\frac{-239-2\sqrt{26273}}{231},\tag 8.19$$
with respective eigenvectors
$$v_1=\bmatrix \displaystyle\frac{16}{15}\\
\noalign{\medskip}
\displaystyle\frac{62}{15}\\
\noalign{\medskip}
1
\endbmatrix ,\quad
v_2=\bmatrix \displaystyle\frac{-73+\sqrt{26273}}{102}\\
\noalign{\medskip}
\displaystyle\frac{2399+3\sqrt{26273}}{1054}\\
\noalign{\medskip}
1
\endbmatrix ,\quad
v_3=\bmatrix \displaystyle\frac{-73-\sqrt{26273}}{102}\\
\noalign{\medskip}
\displaystyle\frac{2399-3\sqrt{26273}}{1054}\\
\noalign{\medskip}
1
\endbmatrix .\tag 8.20$$
We form the columns of the constant matrix $\Cal S$ appearing in (7.17) by using (8.20) to
have
$$\Cal S=\bmatrix
\displaystyle\frac{16}{15}&  \displaystyle\frac{-73+\sqrt{26273}}{102}&\displaystyle\frac{-73-\sqrt{26273}}{102}\\
\noalign{\medskip}
\displaystyle\frac{62}{15}&\displaystyle\frac{2399+3\sqrt{26273}}{1054}& \displaystyle\frac{2399-3\sqrt{26273}}{1054}\\
\noalign{\medskip}
1&1&1
\endbmatrix .\tag 8.21$$
Using (8.18) and (8.21) in (7.17) we see that
$$P_1=I_3,\quad P_2=I_3.\tag 8.22$$

Note that the determinant of $f(0,0)$ given in (8.16) is $25/77$ and hence
not zero, and thus we can choose the
constant $a$ appearing in (2.23) as zero.
Then, using (8.18) and (8.21)
in (2.38) and (2.39) we obtain
$$\omega_1(0)=\bmatrix 0&0&0\\
\noalign{\medskip}
0&0&0\\
\noalign{\medskip}
0&0&0\endbmatrix
 ,\quad \omega'_1(0)=\bmatrix 0&0&0\\
\noalign{\medskip}
0&0&0\\
\noalign{\medskip}
0&0&0\endbmatrix
.\tag 8.23$$
Similarly, using (8.3) in (3.26) we obtain
$$q_1(0)=\bmatrix \displaystyle\frac{16}{15}&0&0\\
\noalign{\medskip}
0&-\displaystyle\frac{2}{5}&\displaystyle\frac{2438}{385}\\
\noalign{\medskip}
0&\displaystyle\frac{2438}{385}&-\displaystyle\frac{275162}{29645}\endbmatrix
.\tag 8.24$$
 From (2.45) and (8.2) we obtain
$$\varphi(0,0)=\bmatrix 0&0&1\\
\noalign{\medskip}
0&0&1\\
\noalign{\medskip}
0&0&1\endbmatrix.\tag 8.25$$
Then, using (8.2), (8.23)-(8.25) in (7.11) and (7.12) we obtain
$$\Cal R=\bmatrix 0&0&\displaystyle\frac{3}{5}\\
\noalign{\medskip}
0&0&-\displaystyle\frac{31}{15}\\
\noalign{\medskip}
0&0&\displaystyle\frac{77}{15}\endbmatrix
,\quad F_2=
\bmatrix 0&0&-\displaystyle\frac{16}{25}\\
\noalign{\medskip}
0&0&-\displaystyle\frac{100}{3}\\
\noalign{\medskip}
0&0&\displaystyle\frac{70148}{1155}\endbmatrix
.\tag 8.26$$
Next, using (8.18), (8.19), (8.21), and (8.22) in (7.17) and (7.18) we get
$$\Cal D_0=\bmatrix
\displaystyle\frac{-239+2\sqrt{26273}}{231}&0\\
\noalign{\medskip}
0&\displaystyle\frac{-239-2\sqrt{26273}}{231}\endbmatrix
,\tag 8.27$$
and using (8.21), (8.22), and (8.26) in (7.19) and (7.20) we obtain
$$\Cal A_1=\bmatrix \displaystyle\frac{6971 i}{623}\endbmatrix,\quad
\Cal B_1=\bmatrix \displaystyle\frac{6971 i}{623}&  \displaystyle\frac{6971 i}{623}\endbmatrix
,\quad \Cal C_1=\bmatrix -\displaystyle\frac{76268 i}{9345}-\displaystyle\frac{11541857 i}{9345 \sqrt{26273}}\\
\noalign{\medskip}  -\displaystyle\frac{76268 i}{9345}+\displaystyle\frac{11541857 i}{9345 \sqrt{26273}}\endbmatrix ,
 \tag 8.28$$
 $$
 \Cal D_1=\bmatrix -\displaystyle\frac{76268 i}{9345}-\displaystyle\frac{11541857 i}{9345 \sqrt{26273}}
 &  -\displaystyle\frac{76268 i}{9345}-\displaystyle\frac{11541857 i}{9345 \sqrt{26273}}\\
\noalign{\medskip}  -\displaystyle\frac{76268 i}{9345}+\displaystyle\frac{11541857 i}{9345 \sqrt{26273}}&
 -\displaystyle\frac{76268 i}{9345}+\displaystyle\frac{11541857 i}{9345 \sqrt{26273}}
\endbmatrix ,
 \tag 8.29$$
 $$\Cal A_2=\bmatrix -\displaystyle\frac{427808}{3115}\endbmatrix,\quad
\Cal B_2=\bmatrix -\displaystyle\frac{427808}{3115}&  -\displaystyle\frac{427808}{3115}\endbmatrix ,
\tag 8.30$$
$$\Cal C_2=\bmatrix \displaystyle\frac{10180418}{102795}+\displaystyle\frac{7539081034}{513975 \sqrt{26273}} \\
\noalign{\medskip}  \displaystyle\frac{10180418}{102795}-\displaystyle\frac{7539081034}{513975 \sqrt{26273}}\endbmatrix ,
 \tag 8.31$$
 $$
 \Cal D_2=\bmatrix \displaystyle\frac{10180418}{102795}+\displaystyle\frac{7539081034}{513975 \sqrt{26273}}
 & \displaystyle\frac{10180418}{102795}+\displaystyle\frac{7539081034}{513975 \sqrt{26273}}\\
\noalign{\medskip}  \displaystyle\frac{10180418}{102795}-\displaystyle\frac{7539081034}{513975 \sqrt{26273}}&
 \displaystyle\frac{10180418}{102795}-\displaystyle\frac{7539081034}{513975 \sqrt{26273}}
\endbmatrix .
 \tag 8.32$$
Thus, we have all the ingredients for the verification of the results presented in
Theorem~7.6. Using (8.21), (8.22), (8.26)-(8.32) in (7.31), (7.32),
(7.34), (7.36)-(7.39) we obtain
$$
J(0)=\bmatrix -\displaystyle\frac{5}{3}&0&\displaystyle\frac{16}{9}\\
\noalign{\medskip} 2&-1&2\\
\noalign{\medskip}
1&-\displaystyle\frac{31}{77}&\displaystyle\frac{46}{77}\endbmatrix ,\quad
\dot J(0)=\bmatrix -\displaystyle\frac{2i}{3}&0&\displaystyle\frac{i}{9}\\
\noalign{\medskip} i&0&i\\
\noalign{\medskip}
i&-\displaystyle\frac{108i}{77}&-\displaystyle\frac{31i}{77}\endbmatrix ,\tag 8.33$$
$$\Cal E_1=\bmatrix -\displaystyle\frac{16095714}{48594841}&\displaystyle\frac{22880111}{145784523}&\displaystyle\frac{32930051}{145784523}\\
\noalign{\medskip} -\displaystyle\frac{10281837}{48594841}&-\displaystyle\frac{30462632}{145784523}&\displaystyle\frac{61380319}{145784523}\\
\noalign{\medskip}
\displaystyle\frac{11927250}{48594841}&\displaystyle\frac{8244046}{48594841}&\displaystyle\frac{7573258}{48594841}\endbmatrix ,\tag 8.34$$
$$\Cal E_2=\bmatrix -\displaystyle\frac{855616 i}{34855}&0&0\\
\noalign{\medskip} \displaystyle\frac{855616 i\left( 2003789164+11541857\,\sqrt{26273} \right)}{95754919319475}&0&0\\
\noalign{\medskip}
\displaystyle\frac{855616 i\left( 2003789164-11541857\sqrt{26273} \right)}{95754919319475}&0&0\endbmatrix ,\tag 8.35$$
$$\Cal E_3=\bmatrix \displaystyle\frac{427808 i\left( 2003789164+11541857\sqrt{26273} \right)}{95754919319475}\\
\noalign{\medskip}
\displaystyle\frac{427808 i\left( 2003789164-11541857\,\sqrt{26273} \right)}{95754919319475}\endbmatrix ,
\tag 8.36$$
$$S(0)=\bmatrix -\displaystyle\frac{6809}{6971}&-\displaystyle\frac{558}{6971}&\displaystyle\frac{1386}{6971}\\
\noalign{\medskip}
-\displaystyle\frac{558}{6971}&-\displaystyle\frac{5049}{6971}&-\displaystyle\frac{4774}{6971}\\
\noalign{\medskip}
\displaystyle\frac{1386}{6971}&-\displaystyle\frac{4774}{6971}&\displaystyle\frac{4887}{6971}\endbmatrix ,$$
$$\dot S(0)=\bmatrix \displaystyle\frac{24111452 i}{48594841}&
-\displaystyle\frac{8336928 i}{48594841}&-\displaystyle\frac{12952632 i}{48594841}\\
\noalign{\medskip}
-\displaystyle\frac{8336928 i}{48594841}&\displaystyle\frac{95224498 i}{145784523}&\displaystyle\frac{111299650 i}{145784523}\\
\noalign{\medskip}
-\displaystyle\frac{12952632 i}{48594841}&\displaystyle\frac{111299650 i}{145784523}
&-\displaystyle\frac{124617494 i}{145784523}\endbmatrix .\tag 8.37$$
Finally, by using (8.33)-(8.37), we evaluate (7.30), (7.33), and (7.35)
and obtain
the small-$k$ expansions for
$J(k),$
$J(k)^{-1},$ and
$S(k),$ respectively, which all agree
with the expansions given in (8.6), (8.8), and (8.9).

\vskip 10 pt

\noindent {\bf Acknowledgments.} The research leading to this
article was supported in part by Consejo Nacional de Ciencia y
Tecnolog\'{\i}a (CONACYT) under project CB2008-99100-F
and by the Department of Defense
under grant number DOD-BC063989.

\vskip 10 pt

\noindent {\bf{References}}

\vskip 3 pt

\item{[1]} Z. S. Agranovich and V. A. Marchenko, {\it The inverse problem of
scattering theory,} Gordon and Breach, New York, 1963.

\item{[2]} T. Aktosun,
{\it Factorization and small-energy asymptotics for the radial Schr\"odinger equation,}
J. Math. Phys. {\bf 41}, 4262--4270 (2000).

\item{[3]} T. Aktosun and M. Klaus,
{\it Small-energy asymptotics for the Schr\"odinger equation on the line,}
Inverse Problems {\bf 17}, 619--632 (2001).

\item{[4]} T. Aktosun, M. Klaus, and C. van der Mee,
{\it Small-energy asymptotics of the scattering matrix for the matrix
Schr\"odinger equation on the line,}
J. Math. Phys. {\bf 42}, 4627--4652 (2001).

\item{[5]} T. Aktosun, M. Klaus, and R. Weder,
{\it Small-energy analysis for the self-adjoint matrix
Schr\"odinger operator on the half line,}
J. Math. Phys. {\bf 52}, 102101 (2011).

\item{[6]} T. Aktosun and R. Weder,
{\it High-energy analysis and Levinson's theorem for the self-adjoint matrix Schr\"odinger operator on the half line,}
J. Math. Phys. {\bf 54}, 012108 (2013).

\item{[7]} G. Berkolaiko, R. Carlson, S. A. Fulling, and P. Kuchment (eds.),
{\it Quantum graphs and their applications,} Contemporary Mathematics, 415,
Amer. Math. Soc., Providence, RI, 2006.

\item{[8]} J. Boman and P. Kurasov,
{\it Symmetries of quantum graphs and the inverse scattering problem,}
Adv. Appl. Math. {\bf 35}, 58--70 (2005).

\item{[9]} E. A. Coddington and N. Levinson, {\it Theory of differential equations,} New York, McGraw-Hill, 1955.

\item{[10]} P. Deift and E. Trubowitz, {\it Inverse scattering
on the line,} Commun. Pure Appl. Math. {\bf 32}, 121--251 (1979).

\item{[11]} P. Exner, J. P. Keating, P. Kuchment, T. Sunada, and A. Teplyaev (eds.),
{\it Analysis on graphs and its applications,}
Proc. Symposia in Pure Mathematics, 77,
Amer. Math. Soc., Providence, RI, 2008.

\item{[12]} L. D. Faddeev, {\it
Properties of the $S$-matrix of the one-dimensional Schr\"odinger equation,} Amer. Math. Soc. Transl. {\bf 65} (ser. 2), 139--166 (1967).

\item{[13]} N. I. Gerasimenko,
{\it The inverse scattering problem on a noncompact graph,}
Theoret. Math. Phys. {\bf 75}, 460--470 (1988).

\item{[14]} N. I. Gerasimenko and B. S. Pavlov,
{\it A scattering problem on noncompact graphs,}
Theoret. Math. Phys. {\bf 74}, 230--240 (1988).

\item{[15]} B. Gutkin and U. Smilansky,
{\it Can one hear the shape of a graph?}
J. Phys. A {\bf 34}, 6061--6068 (2001).

\item{[16]} M. S. Harmer, {\it Inverse scattering for the matrix Schr\"odinger
operator and Schr\"odinger operator on
graphs with general self-adjoint boundary conditions,}
ANZIAM J. {\bf 44}, 161--168 (2002).

\item{[17]} M. S. Harmer, {\it The matrix Schr\"odinger operator
and Schr\"odinger operator on graphs,} Ph.D. thesis, University of
Auckland, New Zealand, 2004.

\item{[18]} M. Harmer, {\it Inverse scattering on matrices with boundary conditions,}
J. Phys. A {\bf 38}, 4875--4885 (2005).

\item{[19]} M. Klaus, {\it
Low-energy behaviour of the scattering matrix for the Schr\"odinger equation on the line,} Inverse Problems {\bf 4}, 505--512 (1988).

\item{[20]} M. Klaus, {\it Exact behavior of Jost functions at low energy,}
J. Math. Phys. {\bf 29}, 148--154 (1988).

\item{[21]} V. Kostrykin and R. Schrader,
{\it Kirchhoff's rule for quantum wires,} J. Phys. A {\bf 32}, 595--630
(1999).

\item{[22]} V. Kostrykin and R. Schrader,
{\it Kirchhoff's rule for quantum wires. II: The inverse problem with possible applications to quantum computers,} Fortschr. Phys. {\bf 48}, 703--716
(2000).

\item{[23]} P. Kuchment,
{\it Quantum graphs. I. Some basic structures,}
Waves Random Media {\bf 14}, S107--S128 (2004).

\item{[24]} P. Kuchment,
{\it Quantum graphs. II. Some spectral properties of quantum and combinatorial graphs,}
J. Phys. A {\bf 38}, 4887--4900 (2005).

\item{[25]} P. Kurasov and M. Nowaczyk,
{\it Inverse spectral problem for quantum graphs,}
J. Phys. A {\bf 38}, 4901--4915 (2005).

\item{[26]} P. Kurasov and F. Stenberg,
{\it On the inverse scattering problem on branching graphs,}
J. Phys. A {\bf 35}, 101--121 (2002).

\end